%% file: main.tex
\PassOptionsToPackage{dvipsnames}{xcolor}
\PassOptionsToClass{nonacm}{acmart}

\documentclass[sigplan,10pt]{acmart}

\input{settings}

\setcopyright{none}
\authorsaddresses{}
\renewcommand{\email}[2][]{}

\begin{document}

\title{\sys{}: Dynamic Detection of Race Conditions in On-Device Network Operating Systems}

\author{Runze Wu}
\affiliation{%
  \department{SKL of Novel Soft. Tech.}
  \institution{Nanjing University}
  \city{Nanjing}
  \country{China}}
\email{runzewu@smail.nju.edu.cn}

\author{Jingbo Zhai}
\affiliation{%
  \department{SKL of Novel Soft. Tech.}
  \institution{Nanjing University}
  \city{Nanjing}
  \country{China}}
\email{jingbozhai@smail.nju.edu.cn}

\author{Shanming Ping}
\affiliation{%
  \institution{Huawei Technologies Co., Ltd}
  \city{Beijing}
  \country{China}}
\email{pingshanming@huawei.com}

\author{Lingzhi Ouyang}
\affiliation{%
  \institution{Huawei Technologies Co., Ltd}
  \city{Shenzhen}
  \country{China}}
\email{ouyanglingzhi1@huawei.com}

\author{Hua Duan}
\affiliation{%
  \institution{Huawei Technologies Co., Ltd}
  \city{Hangzhou}
  \country{China}}
\email{rancle\_dh@163.com}

\author{Qin Zou}
\affiliation{%
  \institution{Huawei Technologies Co., Ltd}
  \city{Hangzhou}
  \country{China}}
\email{zouqin@huawei.com}

\author{Chengcheng Huang}
\affiliation{%
  \institution{Huawei Technologies Co., Ltd}
  \city{Hangzhou}
  \country{China}}
\email{huangchengcheng@huawei.com}

\author{Bingshe Liu}
\affiliation{%
  \institution{Huawei Technologies Co., Ltd}
  \city{Beijing}
  \country{China}}
\email{liubingshe@huawei.com}

\author{Xudong Lang}
\affiliation{%
  \institution{Huawei Technologies Co., Ltd}
  \city{Beijing}
  \country{China}}
\email{langxudong@huawei.com}

\author{Xiaoxing Ma}
\affiliation{%
  \department{SKL of Novel Soft. Tech.}
  \institution{Nanjing University}
  \city{Nanjing}
  \country{China}}
\email{xxm@nju.edu.cn}

\author{Yu Huang}
\authornote{Corresponding author.}
\affiliation{%
  \department{SKL of Novel Soft. Tech.}
  \institution{Nanjing University}
  \city{Nanjing}
  \country{China}}
\email{yuhuang@nju.edu.cn}

\renewcommand{\shortauthors}{R. Wu, J. Zhai, S. Ping, L. Ouyang, H. Duan, Q. Zou, C. Huang, B. Liu, X. Lang, X. Ma, and Y. Huang}


\pagestyle{standardpagestyle}

\input{abs}

\maketitle

\input{intro}
\input{model}
\input{sys}

\input{impl}
\input{exp}
\input{rw}
\input{conclusion}

\begin{acks}
We thank the anonymous reviewers and our shepherd for their insightful comments and suggestions.
This work is supported by the Frontier Technologies R\&D Program of Jiangsu (BF2024059),
the National Natural Science Foundation of China (62372222), 
the CCF-Huawei Populus Grove Fund (CCF-HuaweiFM202505), and    
``111 Center'' (No.~B26023).

\end{acks}

\bibliographystyle{ACM-Reference-Format}
\bibliography{main}

\clearpage
\appendix
\emergencystretch=1em
\input{app-testing-factory}
\input{app-orchestration}
\input{app-implementation}

\end{document}

%% file: settings.tex
\usepackage{tikz}
\usepackage{pgfplots}
\pgfplotsset{compat=1.18}
\usepackage{tabularx}
\usetikzlibrary{arrows.meta,positioning,backgrounds,fit,calc,tikzmark,decorations.markings}
\usepackage{xcolor}
\usepackage{listings}
\usepackage{enumitem}
\usepackage{placeins}

\newcommand{\sys}{\textsc{NosRacer}}
\newcommand{\figref}[1]{Figure~\ref{#1}}
\newcommand{\tabref}[1]{Table~\ref{#1}}
\newcommand{\para}[1]{\par\noindent\textbf{#1.}~}
\lstdefinestyle{profileyaml}{
  basicstyle=\small\ttfamily,
  backgroundcolor=\color{white},
  frame=l,
  framerule=0.25pt,
  framesep=0.45em,
  rulecolor=\color{black!28},
  xleftmargin=0.2em,
  xrightmargin=0.2em,
  framexleftmargin=0.2em,
  framexrightmargin=0em,
  framextopmargin=0em,
  framexbottommargin=0em,
  aboveskip=0.2em,
  belowskip=0.05em,
  columns=fullflexible,
  keepspaces=true,
  breaklines=true,
  showstringspaces=false,
  alsoletter={_:-},
  morekeywords={
    test_case:,topology:,class:,duts:,traffic_endpoints:,virtual_links:,
    workload:,update_scope:,steps:,profile:,last_profiled:,
    functional_labels:,nos_function:,service_category:,config_behavior:,
    execution_history:,recent_runs:,median_duration:,function_coverage:,
    ARP:,MGR-A:
  },
  keywordstyle=\color{RoyalBlue!75!black},
  morestring=[b]",
  stringstyle=\color{ForestGreen!45!black},
}

%% file: abs.tex
\begin{abstract}

Commercial on-device network operating systems (NOSes) run
    {the} complex control planes of production routers and switches,
    where configuration update tasks are executed 
    by multiple loosely coupled components 
    through asynchronous message passing. 
Such executions are prone to race conditions: 
    the same ordered input commands may produce 
    different outcomes when messages are delivered 
    in different orders. 
The race conditions are difficult to expose because 
    they often manifest only as subtle, delayed malfunctions. 
Existing static analysis techniques lack scalability and precision 
    for large-scale industrial NOSes, while 
    dynamic ones incur substantial system-execution cost
    when attempting to cover the space of asynchronous message interleavings.

To address the challenges above, we present \sys{}, 
    a dynamic analysis framework for race condition detection %
    in industrial-grade on-device NOSes. 
\sys{} uses a two-phase design. 
The concentration phase reduces analysis scope 
    by leveraging the locality of configuration update tasks.
Race condition detection is then limited to a small subset of 
    involved components and crucial state variables, 
    thereby reducing the detection cost.
The perturbation phase 
    proactively perturbs message asynchrony to 
    increase the likelihood of executions 
    with race-condition-exposing message interleavings.
It keeps perturbation practical 
    by grouping compatible perturbations
    for parallel execution 
    and adaptively strengthening perturbations.

We implement \sys{} in a commercial, actively developed NOS.
\sys{} is integrated into the existing testing factory and
    used to detect race conditions in 5 key control-plane components.
\sys{} achieves 66\% precision and 
    detects 21 race conditions 
    confirmed by developers as severe bugs, 
    while keeping the detection overhead below 10\%.

\end{abstract}

%% file: intro.tex
\relax
\section{Introduction} \label{sec:intro}
\relax

Commercial on-device Network Operating Systems (NOSes), 
    such as Cisco IOS XR, Juniper JunOS, Huawei VRP, and SONiC, 
    are complex software stacks 
    deployed on production routers and switches \cite{ciscoiosxrmodular,junos_overview,huaweivrp8intro,sonichome}.
In a typical NOS control plane, protocol and management functions 
    are implemented by tens to hundreds of loosely coupled components, 
    which coordinate with one another through asynchronous messages \cite{junosarchitecture,huaweivrp8arch,sonicarchitecture}.

\begin{figure}[t]
    \centering
    \resizebox{\columnwidth}{!}{\input{framework-narrowing-solution}}%
    \par\vspace{-0.2cm}
    \caption{A configuration update task.}
    \Description{The top part shows many components of the NOS control plane, but one configuration update task involves only a few of them. In this example, the upstream component C7 sends the commands, and the target component ARP works with the interacting components VLAN and MGR to process them. The bottom part zooms into ARP and shows two of its handlers accessing the crucial variable vlan.status concurrently, while arp.timer is marked as non-crucial.}
    \label{fig:task}%
    \par\vspace{-0.2cm}
\end{figure}
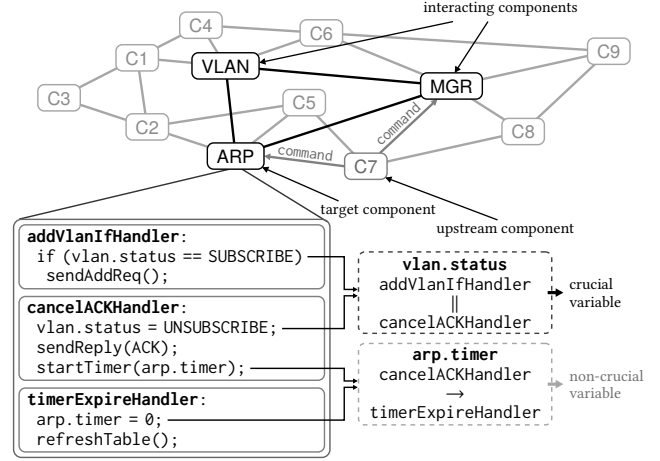

Configuration updates are 
    routine operations in production NOSes,
    e.g., due to topology changes, traffic dynamics, and policy revisions \cite{kim2011evolution,schneider2021snowcap,yuan2025hoyan}.
As shown in \figref{fig:task}, one configuration update task typically 
    consists of an ordered sequence of configuration commands. 
These commands are processed by multiple control-plane components, 
    which interact through asynchronous messages, 
    in order to validate, propagate, and realize the requested state changes.

Correct execution of the configuration update task 
    is of vital importance to the NOS control plane \cite{kakarla2020selfstarter,tang2021campion,tang2023lightyear}.
However, ensuring correct execution of the task 
    is greatly hampered by the \textit{race condition} problem \cite{netzer1992race}. %
That is, in different executions of the same update task, 
    asynchronous message passing
    may result in 
    different ways of coordination among components,
    thus producing 
    different update results.
Race conditions typically do not manifest immediately 
    as user-visible system anomalies.
Instead, they manifest themselves 
    in silent and subtle malfunctions, 
    e.g., a subscription quietly lost, 
    a state left stale, or 
    a downstream notification missed \cite{tang2021campion,lou2022oathkeeper,lou2025t2c}.
By the time severe network failures occur, 
    it is usually too late to identify
    the underlying
    race conditions or trace failures back to them.
This also explains the difficulties in 
    reproducing the race conditions.

\begin{figure}[t]
    \centering
    \resizebox{0.8\columnwidth}{!}{\input{buggy-arp-normal}}%
    \par\vspace{-0.2cm}
    \caption{An exemplar race condition.}
    \Description{The figure shows two possible executions of the same commands from the upstream component, drawn as message sequences between ARP in the middle and MGR on each side. In Possible Execution 1 on the left, ARP ends subscribed. In Possible Execution 2 on the right, the ACK of unsub\_vlan is delayed, so ARP ignores the sub\_vlan command and ends unsubscribed.}
    \label{fig:intro-motivation}%
    \par\vspace{-0.25cm}
\end{figure}
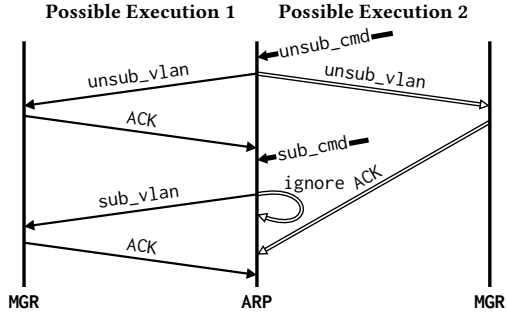

An exemplar race condition from the NOS we study 
    is shown in \figref{fig:intro-motivation}.
An upstream component %
    sends an \textsf{unsub\_vlan} command followed by a \textsf{sub\_vlan} command 
    for the same virtual LAN (VLAN) interface to ARP.
The ARP component needs to interact with the 
    interface management (MGR) component to 
    cancel and reestablish the VLAN subscription.
\begin{itemize}[leftmargin=1em]
    \item \textit{Possible Execution 1}. When the \textsf{ACK} of 
    \textsf{unsub\_vlan} is processed before ARP 
    handles the second \textsf{sub\_vlan} command, 
    ARP observes the updated \textit{unsubscribed} state. 
    It then sends a new \textsf{sub\_vlan} request, 
    and ends \textit{subscribed}, 
    which is the intended outcome. %
    \item \textit{Possible Execution 2}. When the \textsf{ACK} of
    \textsf{unsub\_vlan} is delayed, ARP handles the \textsf{sub\_vlan} command
    first, reads a stale \textit{subscribed} state, and thus ignores the second command.
    ARP ends \textit{unsubscribed}, which is a confirmed bug.
\end{itemize}

Although the sequence of input commands 
    from the upstream component is unchanged, 
    the update of the VLAN status differs
    when the processing order of the two messages differs.
This race condition manifests itself in severe system failures,
    where the VLAN interface cannot forward packets.
Race conditions are notoriously difficult 
    to detect and fix in various 
    concurrent and distributed systems \cite{musuvathi2008chess,kasikci2012portend,liu2017dcatch,cao2026themis}.
In NOSes, each component undergoes 
    extensive unit, integration, and stress testing.
{In the NOS we study,
    a large-scale testing platform called the \textit{testing factory}
    runs these tests.}
{However,} these tests mainly validate 
    component-wise functional
    correctness 
    \cite{xu2021yardstick,singha2024messi,sun2022sieve}.
Race-condition checking is still insufficient.
We analyzed the anomalies observed 
    in a large-scale production NOS deployment (2021\textasciitilde 2024).
Among the 53 issues classified as ``message timing related'',
    27 were caused by race conditions,
    as confirmed by developers.

The race condition problem has been extensively studied
    in concurrent programming \cite{engler2003racerx,flanagan2009fasttrack,roemer2020smarttrack} and 
    event-driven programming \cite{raychev2013effective,hsiao2014cafa,maiya2014race,liu2021threads}.
Both static \cite{blackshear2018racerd,holter2025sound} 
    and dynamic \cite{roemer2020smarttrack,zhang2025efficient} analysis techniques 
    have been developed.
In this work, we study how to cope with race conditions 
    in a production NOS with 200+ control-plane components,
    each typically containing 5k+ variables and 10k+ functions.
At this scale, existing static analysis techniques
    face severe scalability challenges
    due to the sheer number of message interleavings.
Heavy use of indirect calls and asynchronous callbacks in NOS code
    further limits analysis precision
    \cite{li2024lrminer,holter2025sound}.

Although dynamic analysis techniques achieve higher precision
    using concrete execution information,
    they incur unacceptable execution cost
    due to the enormous number of asynchronous executions
    \cite{musuvathi2008chess,leesatapornwongsa2016samc,pick2023psym}.
{Moreover, these techniques typically instrument
    the inter-component messaging service
    to monitor and control message delivery
    \cite{yang2009modist,leesatapornwongsa2016samc,liu2017dcatch,lu2018cloudraid,lu2022cloudraid,tang2024sandtable}.
Since every component communicates through this service,
    such instrumentation affects the entire system
    and is rarely permitted in commercial NOS development.}

To make race condition detection practical
    in a production NOS,
    we propose the dynamic analysis framework \sys{}.
The core innovation of \sys{} lies in 
    its two-phase design.
The \textit{concentration} phase 
    substantially reduces 
    the detection cost, while
    the \textit{perturbation} phase improves its effectiveness 
    in exposing race conditions
    {without instrumenting the messaging service},
    as detailed below.

\para{Concentration}
In the concentration phase, the key issue is 
    to cope with the complexity of an industry-grade NOS.
\sys{} first leverages the existing testing factory of the target NOS 
    to run a small subset of closely related test cases
    and concentrate on a small subset of 
    components.
This is motivated by the observation that 
    configuration update tasks exhibit locality, 
    i.e., each task typically involves only 
    a limited number of components. 
Among the 27 race-condition-induced bugs, every task involves no more than 8 of the 200+ components, and over 80\% involve no more than 5.

After identifying the components 
    involved in a configuration update task,
    \sys{} then leverages static analysis to 
    concentrate on a small subset of crucial variables 
    in component implementations.
This is based on the observation that, in essence, 
    the configuration update is the update of 
    the crucial state information of the involved components.
Such state is typically stored in 
    a small set of crucial variables 
    in the component implementations.
Tracking reads and updates 
    on the variable values 
    can help us efficiently detect race conditions.
That is, if the crucial state information variables 
    are accessed by concurrent operations 
    in asynchronous message handlers, 
    potential race conditions can occur.

\para{Perturbation}
The major obstacle to 
    effective race condition detection in \sys{} 
    is the enormous space of possible system executions 
    induced by asynchronous message interleavings.
As a dynamic analysis tool, \sys{} must 
    carefully plan system executions
    to keep the overheads under control.
To address this obstacle, \sys{} proactively
    perturbs the asynchrony of message passing,
    thereby increasing the likelihood of
    test executions with the specific message interleavings
    needed to expose race conditions.
{\sys{} instruments components
    to inject message delays, leaving the messaging service unchanged.}
In a statistical sense, the perturbation is expected to 
    consistently
    reproduce the race condition
    until it is fixed.
If the update tasks 
    can survive such perturbation, 
    developers can gain more confidence
    in deploying and operating the NOS 
    in production.

The perturbed test execution
    is performed using
    the existing testing factory.
\sys{} consolidates perturbations 
    within the same test execution whenever possible,
    thereby conserving the limited testing budget.
However, two factors fundamentally affect the efficiency of perturbation.

In industrial-grade NOSes, \sys{} lacks fine-grained control
    over message-passing perturbations.
As a result, different perturbations may 
    conflict with one another.
\sys{} groups compatible perturbations 
    within the same test execution and
    dispatches multiple groups of perturbations 
    to the testing factory 
    for parallel execution.

A perturbation may also fail to 
    produce a perturbed execution 
    with the message ordering 
    required for race condition detection.
Based on failed perturbed executions, 
    \sys{} adaptively strengthens the perturbation
    {by increasing the injected delay} to 
    improve the likelihood 
    of successful race condition detection.
Due to the side effects of imprecise perturbation, 
    increasing the perturbation strength is not 
    always beneficial. 
Excessive perturbation may drive 
    the system execution away 
    from the original execution, 
    rendering it ineffective for 
    race condition detection.
\sys{} estimates an upper bound 
    on perturbation strengthening 
    by statically analyzing how timeout mechanisms 
    are implemented in the system.
\sys{} further uses a small number of 
    perturbed executions to 
    empirically estimate this upper bound.

\para{Key results}
{\sys{} has been applied to the monthly release testing of a commercial on-device
    NOS under continuous integration and delivery.
{Over more than a decade of development, this NOS has grown to 200+
    control-plane components and 100M+ lines of code.
These components implement functions such as routing (BGP and OSPF),
    IP services (ARP, DHCP, and VRRP), security (access control and firewalls),
    and device management (CLI and NETCONF).
The NOS runs on routers and switches in carrier backbone and metro networks,
    as well as enterprise campus and data-center networks.
It has been deployed on more than two million devices in over 100 countries.}

The testing factory of this NOS runs system-level tests
    in parallel on a variety of virtualized network topologies.
\sys{} is integrated into this factory and reuses
    its 200,000+ configuration-update test cases.
\sys{} currently covers 5 control-plane components which provide critical network functions
    and have a history of recurring race-condition bugs.
It uncovered 15 previously unknown bugs and reproduced 6 known ones,
    with 66\% precision.
All 21 bugs have been confirmed and fixed by the NOS developers.
Total human effort was at most 20 person-days,
    mostly spent on report analysis,
    and testing cost increased by less than 10\%.}

\para{Summary} 
This work makes three main contributions.
\begin{itemize}[leftmargin=1em,beginpenalty=10000]
    \item We propose \sys{}, a dynamic analysis framework 
        which makes race condition detection practical
        for configuration updates
        in industrial-grade NOSes.
    \item We implement \sys{} in a commercial NOS 
        under active development and 
        integrate it into the existing testing factory.
    \item We apply \sys{} to detect race conditions 
        in 5 components of the NOS under test, and 
        present detailed evaluations 
        showcasing its effectiveness and efficiency.
\end{itemize}

%% file: framework-narrowing-solution.tex
\begin{tikzpicture}[
    remember picture,
    font=\fontfamily{qhv}\selectfont\Large,
    transform shape,
    line width=1.0pt,
    >={Triangle[width=8pt, length=8pt]},
    annotation/.style={font=\rmfamily\fontsize{10}{11}\selectfont},
    message label/.style={font=\ttfamily\fontsize{10}{11}\selectfont,text=black!60},
    callout/.style={line width=0.83877pt,-{Triangle[width=4pt,length=4pt]}},
    title/.style={
        font=\fontfamily{qhv}\selectfont\bfseries\Large
    },
    box/.style={
        rounded corners=4pt,
        draw=black,
        fill=white,
        line width=0.9pt,
        inner xsep=6pt,
        inner ysep=4pt,
        outer sep=0pt,
        align=center,
        font=\fontfamily{qhv}\fontsize{11.5}{12.5}\selectfont
    },
    graybox/.style={
        rounded corners=4pt,
        draw=black!35,
        fill=white,
        line width=0.9pt,
        inner xsep=6pt,
        inner ysep=4pt,
        outer sep=0pt,
        align=center,
        font=\fontfamily{qhv}\fontsize{11.5}{12.5}\selectfont,
        text=black!45
    },
    weakline/.style={
        draw=black!35,
        line cap=round,
        line width=1.5pt
    },
    codebox/.style={
        line width=0.9pt,
        draw=black!50,
        fill=white,
        rounded corners=4pt,
        align=left,
        anchor=north west,
        font=\ttfamily\fontsize{12}{12.5}\selectfont,
        inner xsep=5pt,
        inner ysep=2.5pt,
        text width=6.4cm,
    },
    markline/.style={
    inner sep=0pt,
    outer sep=0pt,
    minimum height=0pt,
    anchor=base
},
]

\pgfdeclarelayer{background}
\pgfsetlayers{background,main}

\begin{scope}[xshift=0.5cm]
\node[box] (VLAN) at (-1.12000, 1.09360) {VLAN};
\node[box] (ARP) at (-0.89600, -0.90000) {ARP};
\node[box] (MGR) at (3.875064, 0.649058) {MGR};
\node[graybox] (C3) at (-4.80000, 0.38800) {C3};
\node[graybox] (C4) at (-1.66400, 1.99200) {C4};
\node[graybox] (C9) at (7.352873, 1.417016) {C9};
\node[graybox] (C8) at (5.459809, -0.357692) {C8};
\node[graybox] (C1) at ($(C3)!0.55081967!(C4)$) {C1};
\node[graybox] (C2) at ($(C3)!0.49963666!(ARP)$) {C2};
\node[graybox] (C6) at ($(C4)!0.2953020134228188!(C9)$) {C6};
\coordinate (taskmid) at ($(VLAN)!0.5!(ARP)$);
\node[graybox] (C5) at ($(taskmid)!0.33333333!(MGR)$) {C5};
\node[graybox, draw=black!50, text=black!60] (C7)
    at ($(MGR)+(C8)-(C9)$) {C7};
\end{scope}

\node[annotation,anchor=north,inner sep=1pt] (targetcomp) at ($(ARP.south east)+(71.92523pt,-21pt)$) {target component};
\draw[callout]([xshift=-25pt,yshift=1pt]targetcomp.north) -- ([xshift=-1.2pt,yshift=1.2pt]ARP.south east);

\node[annotation,anchor=north east] (interacting) at ($(C9.north east)+(-29pt,27pt)$) {interacting components};
\draw[callout]([xshift=-30pt]interacting.south) -- ([yshift=2pt]VLAN.east);
\draw[callout]([xshift=-8pt]interacting.south) -- ([xshift=2pt]MGR.north);

\node[annotation,anchor=north,inner sep=1pt,align=left] (unrelated) at ($(C7.south east)+(74.48598pt,-26.5pt)$) {upstream component};
\draw[callout]([xshift=-5pt,yshift=1pt]unrelated.north) -- ([xshift=-1.2pt,yshift=1.2pt]C7.south east);

\begin{pgfonlayer}{background}
\draw[-{Triangle[width=4pt,length=4pt]}, line width=1.5pt, draw=black!50] (C7.west) -- (ARP.east);
\draw[-{Triangle[width=4pt,length=4pt]}, line width=1.5pt, draw=black!50] (C7) -- (MGR);
\draw[weakline, shorten >=-2pt, shorten <=-2pt] (VLAN) -- (C1);
\draw[weakline, shorten >=-2pt, shorten <=-2pt] (VLAN) -- (C4);
\draw[weakline, shorten >=-2pt, shorten <=-2pt] (VLAN) -- (C6);
\draw[weakline, shorten >=-2pt, shorten <=-2pt] (ARP) -- (C2);
\draw[weakline, shorten >=-2pt, shorten <=-2pt] (ARP) -- (C5);
\draw[weakline, shorten >=-2pt, shorten <=-2pt] (MGR) -- (C6);
\draw[weakline, shorten >=-2pt, shorten <=-2pt] (MGR) -- (C8);
\draw[weakline, shorten >=-2pt, shorten <=-2pt] (MGR) -- (C9);
\draw[weakline, shorten >=-2pt, shorten <=-2pt] (C1) -- (C2);
\draw[weakline, shorten >=-2pt, shorten <=-2pt] (C1) -- (C3);
\draw[weakline, shorten >=-2pt, shorten <=-2pt] (C1) -- (C4);
\draw[weakline, shorten >=-2pt, shorten <=-2pt] (C2) -- (C3);
\draw[weakline, shorten >=-2pt, shorten <=-2pt] (C2) -- (C5);
\draw[weakline, shorten >=-2pt, shorten <=-2pt] (C4) -- (C6);
\draw[weakline, shorten >=-2pt, shorten <=-2pt] (C5) -- (C7);
\draw[weakline, shorten >=-2pt, shorten <=-2pt] (C6) -- (C9);
\draw[weakline, shorten >=-2pt, shorten <=-2pt] (C7) -- (C8);
\draw[weakline, shorten >=-2pt, shorten <=-2pt] (C8) -- (C9);

\draw[weakline, shorten >=-2pt, shorten <=-2pt, draw=black] (ARP) -- (MGR);
\draw[weakline, shorten >=-2pt, shorten <=-2pt, draw=black] (ARP) -- (VLAN);
\draw[weakline, shorten >=-2pt, shorten <=-2pt, draw=black] (VLAN) -- (MGR);
\end{pgfonlayer}

\path (C7.west) -- (ARP.east)
    node[midway, above=1.5pt, sloped, inner sep=0.3pt,
         message label] (cmdarp) {command};
\path (C7) -- (MGR)
    node[pos=0.42, above=2pt, sloped, inner sep=0.3pt,
         message label] (cmdmgr) {command};

\node[codebox] (code1) at (-5.2,-2.405) {
\textbf{addVlanIfHandler}:\\
\hspace*{0.5em}\subnode{line2}{if (vlan.status == SUBSCRIBE)}\\
\hspace*{1em}sendAddReq();\\
};

\node[codebox] (code2) at ($(code1.south west) + (0,-0.14)$) {
\textbf{cancelACKHandler}:\\
\hspace*{0.5em}\subnode{line1}{vlan.status = UNSUBSCRIBE;}\\
\hspace*{0.5em}sendReply(ACK);\\
\hspace*{0.5em}\subnode{line4}{startTimer(arp.timer);}\\
};

\node[codebox] (code3) at ($(code2.south west) + (0,-0.14)$) {
\textbf{timerExpireHandler}:\\
\hspace*{0.5em}\subnode{line3}{arp.timer = 0;}\\
\hspace*{0.5em}refreshTable();\\
};

\node[
    line width=0.9pt,
    draw=black!60,
    rounded corners=5pt,
    fit=(code1)(code2)(code3),
    inner sep=3pt
] (codegroup) {};

\begin{pgfonlayer}{background}
\draw[line width=0.9pt,draw=black!80]
    ([xshift=-4pt]ARP.south)
    -- ([xshift=4pt, yshift=-0.3pt]codegroup.north west);

\draw[line width=0.9pt,draw=black!80]
    ([xshift=4pt]ARP.south)
    -- ([xshift=-2pt, yshift=-2pt]codegroup.north east);
\end{pgfonlayer}

\coordinate (gap12) at ($(code1.south)!0.5!(code2.north)$);
\coordinate (gap23) at ($(code2.south)!0.5!(code3.north)$);

\node[
    line width=0.9pt,
    draw=black!75,
    dashed,
    rounded corners=4pt,
    align=center,
    font=\ttfamily\fontsize{12}{12}\selectfont,
    inner xsep=6pt,
    inner ysep=2pt,
    text width=3.85cm,
    minimum height=1.8cm,
    anchor=west
] (vlanbox) at ($(codegroup.east |- gap12)+(0.6cm,0)$) {\textbf{vlan.status}\\[1pt]
addVlanIfHandler\\
$\parallel$\\
cancelACKHandler};

\node[
    line width=0.9pt,
    draw=black!35,
    dashed,
    rounded corners=4pt,
    align=center,
    font=\ttfamily\fontsize{12}{12}\selectfont,
    inner xsep=6pt,
    inner ysep=2pt,
    text width=3.85cm,
    minimum height=1.8cm,
    anchor=west
] (arpbox) at ($(codegroup.east |- gap23)+(0.6cm,0)$) {\textbf{arp.timer}\\[1pt]
cancelACKHandler\\
$\rightarrow$\\
timerExpireHandler};

\draw[-{Triangle[length=3pt,width=3pt]}, line width=0.9pt]
    ([xshift=-3.0pt]line1.east)
    -| ([xshift=-0.3cm,yshift=-2pt]vlanbox.west)
    -- ([yshift=-2pt]vlanbox.west);
    
\draw[-{Triangle[length=3pt,width=3pt]}, line width=0.9pt]
    ([xshift=-3.0pt]line2.east)
    -| ([xshift=-0.3cm,yshift=2pt]vlanbox.west)
    -- ([yshift=2pt]vlanbox.west);

\draw[-{Triangle[length=3pt,width=3pt]}, line width=0.9pt]
    ([xshift=-3.0pt]line3.east)
    -| ([xshift=-0.3cm,yshift=-2pt]arpbox.west)
    -- ([yshift=-2pt]arpbox.west);

\draw[-{Triangle[length=3pt,width=3pt]}, line width=0.9pt]
    ([xshift=-3.0pt]line4.east)
    -| ([xshift=-0.3cm,yshift=2pt]arpbox.west)
    -- ([yshift=2pt]arpbox.west);

\node[
    annotation,
    align=left,
    anchor=west
] (crucial) at ($(vlanbox.east)+(0.25cm,0)$) {crucial\\ variable};

\draw[-{Triangle[length=5pt,width=5pt]}, line width=2pt]([xshift=-3pt]vlanbox.east) -- ([xshift=7pt]vlanbox.east);

\node[
    annotation,
    align=left,
    text=black!55,
    anchor=west
] (filtered) at ($(arpbox.east)+(0.25cm,0)$) {non-crucial\\ variable};

\draw[-{Triangle[length=5pt,width=5pt]}, line width=2pt, draw=black!35,]([xshift=-3pt]arpbox.east) -- ([xshift=7pt]arpbox.east);

\end{tikzpicture}

%% file: buggy-arp-normal.tex
\begin{tikzpicture}[
    y=0.83cm, %
    font=\ttfamily\Large,
    >={Triangle[width=6pt, length=6pt]},
    transform shape,
    lifeline/.style={
        line width=2pt
    },
    msg/.style={
        ->,
        line width=1.3pt
    },
    msg2/.style={
        ->,
        double,
        double distance=1.0pt,
        line width=0.8pt,
        >={Triangle[open, width=6pt, length=6pt]}
    },
    climsg/.style={
        -{Triangle[width=6pt, length=6pt]},
        line width=3.0pt
    },
    title/.style={
        font=\bfseries\Large
    },
    executiontitle/.style={
        font=\bfseries\Large
    }
]
\def\gap{4.8}

\def\xa{0}
\def\xb{\gap}
\def\xc{2*\gap}

\def\dy{0.8}

\node[executiontitle] at ({(\xa+\xb)/2},0.55) {Possible Execution 1};
\node[executiontitle] at ({(\xb+\xc)/2},0.55) {Possible Execution 2};

\draw[msg] (\xb,-1) -- (\xa,{-1-\dy})
    node[midway, above, sloped] {unsub\_vlan};

\draw[msg] (\xa,-2.05) -- (\xb,{-2.05-\dy})
    node[midway, above, sloped] {ACK};

\draw[msg] (\xb,-4) -- (\xa,{-4-\dy})
    node[midway, above, sloped] {sub\_vlan};

\draw[msg] (\xa,-5.2) -- (\xb,{-5.2-\dy})
    node[midway, above, sloped] {ACK};

\draw[msg2] (\xb,-1) -- (\xc,{-1-\dy})
    node[midway, above, sloped] {unsub\_vlan};

\draw[msg2] (\xc,-2.2) -- (\xb,-5.5)
    node[midway, above, sloped] {ACK};

\draw[msg2](\xb,-4).. controls ({\xb+1.2},-3.7) and ({\xb+1.2},-5.0) ..(\xb,-4.5)
    node[midway, above, xshift=8pt, yshift=5pt] {ignore};

\draw[climsg] ({\xb+2.8},-0.0) -- (\xb,-0.6)
    node[midway, sloped, fill=white, inner sep=0.7pt, yshift=1pt] {unsub\_cmd};

\draw[climsg] ({\xb+2.3},-2.65) -- (\xb,-3.15)
    node[midway, sloped, fill=white, inner sep=0.7pt, yshift=1pt] {sub\_cmd};

\def\lifelinebottom{-6.3}
\draw[lifeline] (\xa,-0.2) -- (\xa,\lifelinebottom);
\draw[lifeline] (\xb,-0.2) -- (\xb,\lifelinebottom);
\draw[lifeline] (\xc,-0.2) -- (\xc,\lifelinebottom);

\node[title, anchor=north, inner ysep=0pt] at (\xa,{\lifelinebottom-0.20}) {\texttt{MGR}};
\node[title, anchor=north, inner ysep=0pt] at (\xb,{\lifelinebottom-0.20}) {\texttt{ARP}};
\node[title, anchor=north, inner ysep=0pt] at (\xc,{\lifelinebottom-0.20}) {\texttt{MGR}};

\end{tikzpicture}

%% file: model.tex
\section{\sys{} Happen-Before Model} \label{sec:modeling}

At the core of race condition detection lies the problem of 
    reasoning about the temporal ordering relations 
    among shared-resource accesses \cite{lamport1978time}.
This section presents the happen-before model
    behind \sys{}.
It first defines the configuration update task, and then 
    the happen-before relation between operations in the task.
The definition of a race condition follows from
    that of the happen-before relation.

\subsection{Configuration Update Task} \label{sec:task}

Configuration updates are 
    routine in production NOSes.
A \textit{configuration update task} consists of 
    a sequence of configuration update commands
    issued {as messages} by upstream components
    in the NOS control plane.
A task is carried out by multiple components 
    that coordinate through asynchronous messages.

Since components interact via message passing,
    component execution follows an event-driven model.
Upon the delivery of one message,
    the component will initiate the corresponding \textit{message handler}
    to process the message.
The message handler consists of two types of operations.
One is local operations, which mainly read and update
    local information of the component.
The other is message-sending operations,
    which result in message handler execution in another component.

In each run, \sys{} selects one task and 
    designates one component involved in the task 
    as the \textit{target component}. 
\sys{} analyzes race conditions
    on a per-component basis because 
    components constitute the natural 
    ownership and concurrency boundaries 
    in the NOS control plane. 
This design improves effectiveness by 
    localizing potential race conditions to 
    the state variables and message-handling logic 
    of individual components.
The detection of race conditions utilizes 
    the asynchronous interaction between 
    the target component and 
    the \textit{interacting components}. 
Instead of reasoning about all possible shared-resource accesses and
    message interleavings at the system level,
    \sys{} restricts the analysis to a small subset of components, %
    thereby reducing the search space. %

The key to race condition detection is 
    the reasoning about 
    the temporal happen-before relations 
    among operations in message handlers, %
    as defined in detail below.

\subsection{Happen-before Model} \label{subsec:happen-before}

The happen-before model primarily defines 
    the temporal relations among operations, 
    which may occur within the same component or 
    across different components.
Operations within the same component are totally ordered
    by the serial execution
    of its message handlers.
Operations on different components are ordered 
    by the causality inherent in message passing.
We define these ordering relations as follows.

\para{Handler program order}
The operations inside one message handler 
    are totally ordered. %
Specifically, for any two operations $op_1$ and $op_2$ 
    in one message handler, %
    if $op_1$ is executed before $op_2$,
    we have: $op_1\overset{P}{\to}op_2$. %

\para{Message passing order}
The message sending operation and 
    the corresponding message handling operations 
    are temporally ordered.
Specifically, assume that operation $op_1$ 
    is a message sending operation 
    in message handler $hdl_1$ 
    in some component $C_1$, 
    which results in the message delivery on another component $C_2$. 
The message delivery is processed 
    by message handler $hdl_2$.
For any operation $op_2 \in hdl_2$, we have: $op_1 \overset{M}{\to} op_2$.

Note that not all operations in $hdl_1$ happen-before operations in $hdl_2$.
Only $op_1$ (as well as operations before $op_1$ in $hdl_1$) happens-before operations in $hdl_2$.
This subtlety will affect our definition of the happen-before relationship between message handlers below.

\para{FIFO channel order}
When two message sending operations in one component 
    have already been ordered based on $\overset{P}{\to}$ and $\overset{M}{\to}$ defined above,
    this order is preserved by the FIFO property
    of the message channel.
Specifically, assume that $snd_1$ and $snd_2$ are 
    two sending operations on the same component,
    and these two messages have the same receiver component.
Let $hdl_1$ and $hdl_2$ denote the message handlers of $snd_1$ and $snd_2$ respectively.
If $snd_1$ precedes $snd_2$ with respect to the transitive closure of $\overset{P}{\to}$ and $\overset{M}{\to}$, we have:
\begin{align*}
    \forall op_1 \in hdl_1, \forall op_2 \in hdl_2, op_1 \overset{F}{\to} op_2
\end{align*}

\para{Task order}
Commands within a task 
    are totally ordered.
When two ordered commands are sent 
    from one upstream component 
    to one (target or interacting) component, 
    this command order is preserved 
    by the FIFO message channel.
Specifically, we have two commands $cmd_1$ and $cmd_2$ 
    from one upstream component and 
    $cmd_1$ precedes $cmd_2$.
If $cmd_1$ and $cmd_2$ result in 
    two message handlers $hdl_1$ and $hdl_2$ 
    on one receiver component, 
    we have:
\begin{align*}
    \forall op_1 \in hdl_1, \forall op_2 \in hdl_2, op_1 \overset{T}{\to} op_2
\end{align*}

\noindent The task order is 
    in essence the same as the FIFO channel order.
The main difference is that the task order 
    originates from the order in the input commands in a given task, 
    while the FIFO channel order originates from 
    the sequential local processing and
    the causality in message passing.
Note that in subsequent race condition checking, 
    the task order serves as the basis for determining 
    whether the external test inputs have changed.

\para{Happen-before order}
The {four} ordering relations defined above
    collectively form the happen-before relation between operations.  
Specifically, for two operations $op_1$ and $op_2$ in one or two components, the happen-before relation $\to$ between $op_1$ and $op_2$ is defined as the transitive closure of $\overset{P}{\to}$, $\overset{M}{\to}$, $\overset{F}{\to}$ and $\overset{T}{\to}$:
\begin{align*}
    \to \ \triangleq (\overset{P}{\to} \cup \overset{M}{\to} \cup \overset{F}{\to} \cup \overset{T}{\to})^+
\end{align*}

\noindent We denote two operations as \textit{concurrent}, if neither happens-before the other. That is: 
\begin{align*}
    op_1 || op_2 \text{ iff } \neg (op_1\to op_2) \wedge \neg (op_2 \to op_1)
\end{align*}

For ease of presentation, 
    we also define the happen-before relation 
    between message handlers.
The happen-before relation between message handlers 
    follows from its counterpart between operations, 
    except for a few subtle differences.
Specifically, 
    for two message handlers $hdl_1$ and $hdl_2$:
\begin{align*}
    hdl_1 \to hdl_2 \text{, iff } \forall op_1 \in hdl_1, \forall op_2 \in hdl_2, op_1\to op_2
\end{align*}

\noindent Here, we overload `$\to$' to also denote the happen-before relation between message handlers.
Note that the universal quantifier `$\forall$' before $op_1$ and $op_2$ is due to the subtlety that one message handler may have only part of its operations happening-before operations in another handler, as discussed in {the} definition of $\overset{M}{\to}$.

\para{Concrete execution order}
Since each component follows serial execution semantics,
    all operations and handlers within the same component
    are necessarily totally ordered 
    in any concrete execution.
We denote this order 
    in a given concrete execution $\sigma$ 
    as $\to_{\sigma}$.
Note that this order pertaining to one concrete execution
    should not be confused with 
    the happen-before order defined above.
The happen-before order $\to$ is preserved 
    across all executions, rather than 
    being valid only for one single concrete execution.

\subsection{Race Condition} \label{sec:race-condition}

In a configuration update task, a race condition \cite{netzer1992race}
    is characterized by the fact that 
    the same task may lead to different outcomes %
    under different message orderings. %

\para{Crucial variable}
For a configuration update task, the core logic of a component 
    is to update the various states it holds 
    throughout the configuration update process.
The intermediate state of a component is maintained 
    through variables that encode the state information.
We call such variables \textit{crucial variables}.
Once we concentrate on crucial variables, 
    the race condition detection problem reduces to 
    determining whether a crucial variable may be concurrently updated.
Concurrent update operations on a crucial variable 
    can potentially be ordered differently across executions, 
    resulting in nondeterministic values of the variable.
    \sys{} leverages the testing factory
    (\S\ref{sec:impl}) to
    run test cases of configuration update tasks.
Then \sys{} checks concurrent reads and updates
    on crucial variables
    in the test executions.

Here, for ease of presentation, we assume that
    crucial variables can be uniquely identified and their occurrences correlated
    in one test execution and across different test executions.
Note that during actual system execution, 
    a class may be instantiated into 
    multiple concrete objects.
Thus the crucial variable, 
    typically defined in some class, 
    will have multiple concrete instances.
Within an execution, accesses to different instances must be distinguished.
\para{Concurrent and perturbed accesses}
We denote the \textit{concurrent access} 
    on crucial variable $var$
    in some original test execution $\sigma_{ori}$
    as $acc_{con} = (var, hdl_1, hdl_2)$, where:
\begin{align*}
    hdl_1 || hdl_2, \ hdl_1\to_{\sigma_{ori}} hdl_2
\end{align*}

\noindent \sys{} perturbs test executions to obtain a \textit{perturbed access} from one concurrent access (as detailed in \S\ref{subsec:perturb}).
A perturbed access corresponds to the original concurrent access in another execution,
    with the order of the two message handlers reversed.
That is, given one original concurrent access $acc_{con}$, the perturbed access in perturbed execution $\sigma_{per}$ is $acc_{per} = (var, hdl_2, hdl_1)$, where $hdl_2\to_{\sigma_{per}} hdl_1$.
The technical details behind
    instance identification and access alignment
    are provided in \S\ref{sec:impl}.

\para{Race condition}
Given a configuration update task, %
    a \textit{race condition} is detected, if 
    concurrent accesses to a crucial variable may 
    render its value nondeterministic.
Specifically, given
    concurrent access $acc_{con}$ and
    its paired perturbed access $acc_{per}$, 
    a race condition is found if:
\begin{itemize}[leftmargin=1em]
    \item the sequence of input commands remains unchanged after perturbation; 
    \item $(hdl_1 || hdl_2) \wedge (hdl_1\to_{\sigma_{ori}} hdl_2) \wedge (hdl_2\to_{\sigma_{per}} hdl_1)$;
    \item $\textsf{val}_{\sigma_{ori}}(var) \neq \textsf{val}_{\sigma_{per}}(var)$.
\end{itemize}

The concurrent and perturbed accesses, 
    as well as the race condition defined above{,}
    are essentially patterns in test executions.
These patterns may have duplicate occurrences both within a single test execution
and across different test executions.
As for two concurrent accesses, if they access the same crucial variable,
    with the same two message handlers in the same order,
    we refer to the two occurrences as duplicates of each other.
Two duplicate concurrent accesses only differ in where they occur in one test execution or across different executions.
Duplicates of perturbed accesses are defined similarly.
Duplicate concurrent accesses and perturbed accesses may also result in duplicate race conditions.
In our terminology, a race condition refers to an equivalence class of duplicate
race-condition occurrences.
For the ease of presentation, we treat it as one single race condition.
Concurrent accesses and perturbed accesses are treated similarly.

\para{Severity of a race condition}
When race conditions are detected, test engineers 
    are primarily concerned with the severity of the race condition, 
    i.e., how the effects of a race condition propagate and are amplified through subsequent component executions.
In this work, the severity assessment (\S \ref{subsec:detection-validation}) primarily focuses on
    whether a race condition affects 
    the subsequent sequences of 
    message-sending operations and message handler function calls.

%% file: sys.tex
\section{\sys{} Framework} \label{sec:framework}

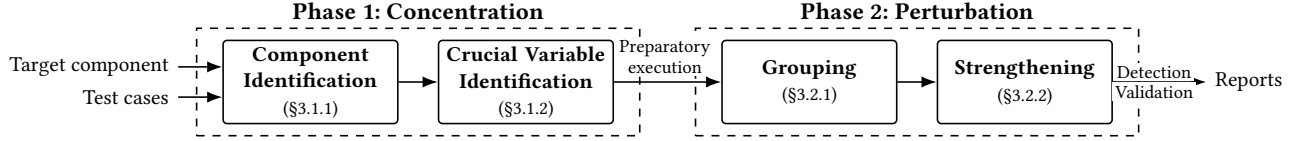
\begin{figure*}[t]
    \centering
    \input{framework-overview}%
    \par\vspace{-0.2cm}
    \caption{The concentration-and-perturbation design of \sys{}.}
    \Description{The figure shows how NosRacer detects race conditions with its two-phase design.}
    \label{fig:phases}%
    \par\vspace{-0.2cm}
\end{figure*}

\sys{} is a dynamic analysis framework
    for race condition detection. %
The practical deployability of \sys{} in industrial-grade NOSes
    stems from two key technical innovations,
    as shown in \figref{fig:phases}.
In the \textit{concentration} phase, \sys{} tackles
    the scalability limitations of existing
    static and dynamic analysis techniques.
In the \textit{perturbation} phase, by amplifying
    the asynchrony of inter-component message interactions,
    \sys{} proactively constructs execution conditions
    that expose race conditions. %

\subsection{Concentration} \label{subsec:concen}

The concentration phase serves two primary purposes:
    reducing hundreds of components to a small subset, and
    thousands of component-state variables to tens or hundreds of crucial ones, %
    as shown in \figref{fig:task}.

\subsubsection{Identification of interacting components} \label{subsubsec:interacting-components}

Our inspection of existing test executions provides the intuition that
    a configuration update task typically involves
    only a small subset of components.
We conducted a comprehensive empirical study
    of the latest bugs related to configuration updates.
The study confirms our intuition.
Among 27 real-world bugs we analyzed,
    all involve no more than 8 interacting components,
    and 22 involve no more than 5.
According to our study, \sys{} efficiently identifies
    the interacting components of the pre-specified target component,
    by running test cases of configuration update tasks. %

Note that static analysis alone
    cannot precisely identify all interacting components
    for a given target component,
    because component interactions are largely
    determined by runtime message routing,
    system state, and asynchronous execution order.
Although static analysis can over-approximate potential interactions,
    such an approximation often includes
    an excessive number of irrelevant components and
    thus provides limited scalability benefits
    in the concentration phase. %

In the testing factory, each test case is annotated
    with rich functional labels.
\sys{} uses the functional labels to select
    test cases pertaining to the target component.
Moreover, the test case profiles provide coverage statistics
    for the target component.
The selected test cases are prioritized
    according to the coverage statistics. %
More information on test case profiles is provided in
\relax
    Appendix~\ref{app:config-update-tests}.%
\relax

Dynamic executions of the selected test cases can
    help \sys{} identify the interacting components
    with respect to the target component.
Specifically, \sys{} constructs a component graph,
    whose nodes are simply the components.
If two components interact with each other via one message,
    a directed edge is added from the message sender to the receiver.
The direct predecessor and direct successor components of the target component
    are identified as interacting components.
Test cases are executed in descending order of priority,
    until the set of interacting components stops changing.
In the practical deployment of \sys{}, this convergence takes no more than
    20 test cases.

\subsubsection{Identification of crucial variables} \label{subsubsec:crucial-variables}
A target component may contain a great number of
    configuration update-related variables
    (typically more than 1,000 in the NOS studied in this work).
However, most of such variables are irrelevant to
    the inter-component message interactions.
\sys{} identifies a variable as crucial,
    when it is accessed by multiple message handlers and
    at least one updates the variable value.
Analysis of crucial variables starts from
    the message handler executions recorded
    in the test executions in the previous step
    (\S \ref{subsubsec:interacting-components}).
\sys{} uses the Abstract Syntax Tree (AST) to
    provide a structured representation
    of the message handling logic,
    allowing \sys{} to identify variable declarations,
    read/update accesses, and
    message passing operations.
By analyzing these syntactic structures
    together with component-level execution contexts,
    \sys{} can statically identify variables
    which are potentially shared across asynchronous executions and
    may therefore give rise to race conditions.
For each identified variable, \sys{} records
    the access sites with their enclosing functions.

Our choice of an AST-based approach is motivated
    by the following subtle tradeoffs.
Compared with Intermediate Representation (IR)-based techniques,
    AST-based analysis preserves source-level program structures and
    can be applied with lower engineering and computational cost,
    making it more scalable for industrial-grade NOS codebases.
It allows \sys{} to identify
    access sites of crucial variables and
    relevant message-handling logic.
However, AST-based analysis provides
    limited semantic information and
    is less precise in reasoning about aliasing,
    inter-procedural data flow and synchronization.
In contrast, IR-based analysis enables
    more precise semantic reasoning,
    but often incurs a higher deployment and analysis cost
    in large production systems.

\subsubsection{Preparatory test executions} \label{subsubsec:preparation}

Given the crucial variables, \sys{} needs
    concrete test executions to determine
    whether the crucial variables are updated
    by concurrent message handlers.
\sys{} leverages coverage information in test case profiles
    to select test cases that cover functions containing
    crucial-variable access sites
    (as recorded in the previous step in \S\ref{subsubsec:crucial-variables}).
\sys{} greedily selects the test case
    that covers the maximum number of uncovered crucial-variable access sites,
    until all the sites are covered.

Given the test executions, \sys{} constructs
    the graph of operations
    in the test execution traces.
{\sys{} pairs sends and receives using message hashes
    recorded at the send/receive APIs.}
The operations are modeled as graph nodes, and
    an edge is added between two operations
    if these two operations are ordered by
    the happen-before relation
    defined in \S \ref{subsec:happen-before}.
Based on the operation graph, \sys{} can
    decide the concurrency between message handlers, and thus can
    detect concurrent accesses on crucial variables.
{This graph enables \sys{} to determine concurrency
    without adding logical timestamps to messages.}
\sys{} uses the concurrent accesses
    to construct perturbed test executions.

\subsection{Perturbation} \label{subsec:perturb}

\begin{figure}[t]
    \centering
    \includegraphics[width=\columnwidth]{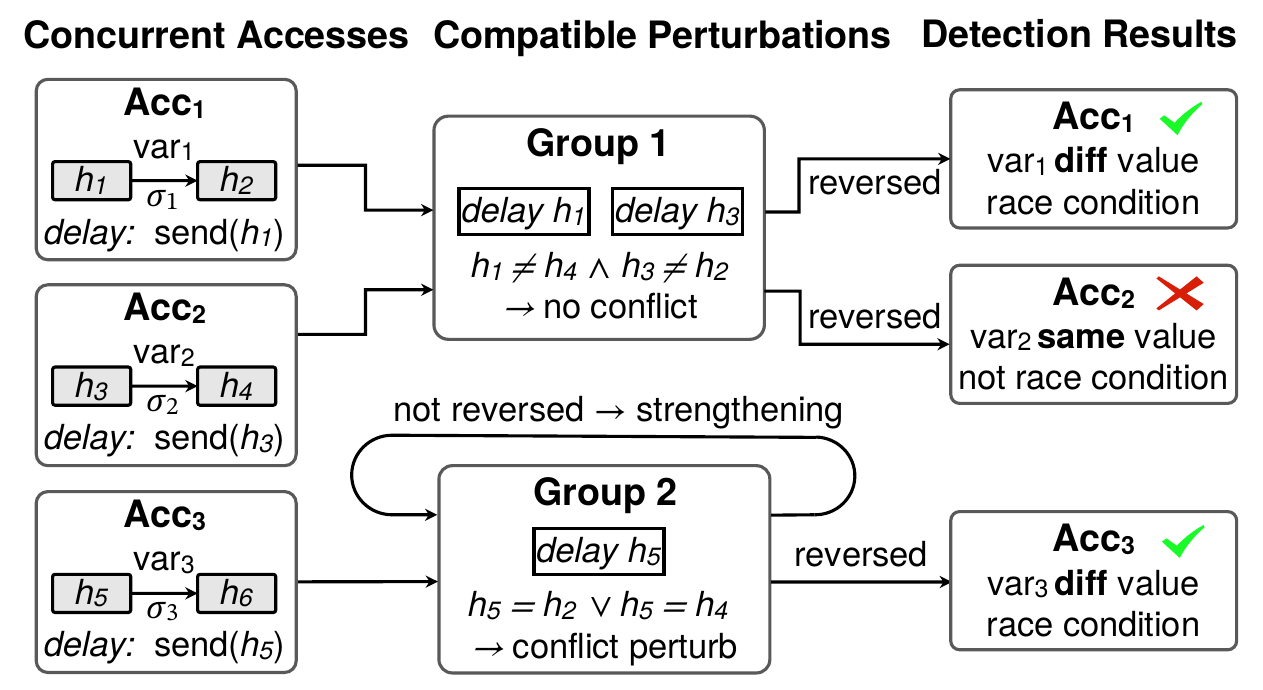}%
    \par\vspace{-0.2cm}
    \caption{Perturbation phase in \sys{}.}
    \Description{The figure shows an example of the perturbation phase.}
    \label{fig:perturbation}%
    \par\vspace{-0.2cm}
\end{figure}

The perturbation phase starts from
    the concurrent accesses on crucial variables
    (\S \ref{subsubsec:preparation}),
    as shown in \figref{fig:perturbation}.
Given concurrent access $acc_{con} = (var, hdl_1, hdl_2)$
    in test execution $\sigma_1$,
    message handlers $hdl_1$ and $hdl_2$
    concurrently access crucial variable $var$
    with respect to the happen-before relation $\to$. %
In the original test execution $\sigma_1$, we have
    $hdl_1 \to_{\sigma_1} hdl_2$.
Therefore, there exists some possible execution $\sigma_2$
    of the same configuration update task,
    in which $hdl_2 \to_{\sigma_2} hdl_1$.
The perturbation phase primarily aims to
    proactively construct $\sigma_2$, a test execution that
    reverses the concrete execution order between $hdl_1$ and $hdl_2$.

The perturbation delays
    the sending operation corresponding to the message handler $hdl_1$
    in the perturbed test execution $\sigma_2$.
The perturbation succeeds if \sys{} finds
    the perturbed access $acc_{per} = (var, hdl_2, hdl_1)$
    in $\sigma_2$,
    corresponding to the original concurrent access $acc_{con}$, where:
\begin{itemize}[leftmargin=1em]
    \item The test inputs (i.e., the task orders among operations)
    remain unchanged from $\sigma_1$ to $\sigma_2$;
    \item $(hdl_1 \to_{\sigma_1} hdl_2) \wedge (hdl_2 \to_{\sigma_2} hdl_1)$.
\end{itemize}

\noindent \sys{} finds a race condition
    if given a successful perturbed access $acc_{per}$,
    $\textsf{val}_{\sigma_1}(var) \neq \textsf{val}_{\sigma_2}(var)$.

The perturbation derived from a concurrent access $acc_{con}$
    is the injection of a certain delay
    before the sending operation that results in $hdl_1$.
\sys{} consolidates perturbations
    within the same test execution whenever possible,
    thereby conserving system execution budget.
At the same time, since perturbations
    targeting different concurrent accesses
    may conflict with one another,
    \sys{} models conflicts among perturbations, and 
    assigns mutually conflicting perturbations
    to separate parallel test executions (\S \ref{subsubsec:grouping-perturbations}). %
Since the perturbations may fail, \sys{} adaptively strengthens the perturbation (\S \ref{subsubsec:strengthening-perturbations}).

\subsubsection{Grouping compatible perturbations} \label{subsubsec:grouping-perturbations}

For concurrent access $acc_{con}$,
    the perturbation delays the sending operation
    corresponding to $hdl_1$.
This delay increases the probability
    that $hdl_2$ accesses $var$ before $hdl_1$. %
Ideally, \sys{} should only delay
    the single message corresponding to $hdl_1$,
    without any side effect on the NOS under test.
However, in production NOSes,
    it is generally difficult to achieve precise delay control
    over specified messages.
The testing environment of \sys{} does not assume
    a centralized overarching scheduler
    capable of controlling nondeterminism at a fine granularity,
    as in distributed system model checking \cite{yang2009modist,leesatapornwongsa2016samc,tang2024sandtable,ouyang2025multi}.
Another promising approach is to hack
    the underlying messaging service
    to gain precise control over
    delays of specific messages.
However, in industrial-grade NOSes,
    the underlying messaging service
    is critical to the development and maintenance
    of the entire system.
As a result, specific testing tools
    are typically not granted the privilege
    to instrument the messaging service.

In practice, when constructing
    happen-before relations among message handlers,
    \sys{} has already established the mapping
    between each message-sending operation and
    its corresponding message handlers.
For concurrent access $acc_{con}$,
    \sys{} locates the message-sending operation
    corresponding to $hdl_1$ and
    instruments the program by
    inserting delay-inducing code
    before that sending operation.

At the source level, \sys{}
    can identify message send sites and
    the types of messages they may send, but
    it cannot precisely distinguish
    individual runtime message instances.
A single source-level send operation
    may be executed multiple times and
    produce multiple messages
    under different runtime contexts.
Therefore, source-level instrumentation
    can only delay a group of messages.
Denote the identifier of a message
    corresponding to message handler $hdl_1$ as
    $\textsf{msgid}(hdl_1) = (\textsf{src}(hdl_1), \textsf{rcv}(hdl_1), \textsf{typ}(hdl_1))$.
Here, $\textsf{src}(hdl_1)$ and $\textsf{rcv}(hdl_1)$ denote
    the message sender and receiver respectively,
    and $\textsf{typ}(hdl_1)$ denotes the message type.
As for concurrent access $acc_{con}$,
    \sys{} injects delay before messages matching %
    $\textsf{msgid}(hdl_1)$. %
Note that the injection of the message delay is not accurate,
    in the sense that all messages of a specific type
    between a specific sender and receiver
    will be delayed.

To save testing budget, \sys{} batches perturbations,
    i.e., injections of message delays,
    given that the perturbations do not counteract each other.
The basic rationale behind the compatibility
    between perturbations
    is as follows.
For two perturbed accesses
    $acc_{per} = (var, hdl_2, hdl_1)$ and $acc_{per}' = (var', hdl_2', hdl_1')$,
    $acc_{per}$ will delay $hdl_2$, while
    $acc_{per}'$ will delay $hdl_2'$.
If $hdl_2' = hdl_1$,
    both $hdl_2$ and $hdl_1$ in $acc_{per}$ will be delayed.
These two delay injections counteract each other,
    reducing the probability that
    the order between $hdl_2$ and $hdl_1$ is reversed.
In this case, the perturbations $acc_{per}$ and $acc_{per}'$
    are deemed \textit{conflicting}, or \textit{not compatible}.

\sys{} partitions the perturbations
    into perturbation groups in which
    every pair of perturbations is compatible.
To execute a group of perturbations,
    \sys{} reruns the selected test cases and
    injects all the delays prescribed in the perturbations.
Different perturbation groups
    are independent.
\sys{} therefore submits
    multiple groups to the testing factory
    for parallel execution.

\subsubsection{Strengthening unsuccessful perturbations} \label{subsubsec:strengthening-perturbations}

The time interval between
    $hdl_1$ and $hdl_2$ in $acc_{con}$ %
    serves as the initial injected delay.
It is an approximate estimation of
    the delay needed to
    reverse $hdl_1$ and $hdl_2$.
Various factors, e.g.,
    different types of message queues widely used in components and
    the message handler execution time,
    can possibly affect whether the order reversal succeeds.
When the perturbation fails,
    if \sys{} can still find the original occurrence
    in the perturbed test execution,
    it simply doubles the delay injection and
    schedules another round of perturbation.

Note that the perturbation process
    typically terminates after only a limited number of rounds, because
    the side effects of perturbation increase significantly
    with the injected delay.
As the delay increases, \sys{} soon
    fails to observe the original concurrent accesses
    in the resulting test executions.
\sys{} statically analyzes the timeout values
    used in the system implementation to estimate
    an upper bound on injected message delays.
\sys{} further uses a small number of perturbed test executions
    to empirically estimate the upper bound
    on injected message delays.
After the perturbation phase, \sys{}
    collects all successfully perturbed accesses
    paired with their original concurrent accesses
    for race condition detection and validation.

\subsection{Detection and Validation} \label{subsec:detection-validation}

\sys{} first performs preparatory test executions to
identify suspected race conditions, i.e., concurrent accesses.
\sys{} then constructs perturbed executions as evidence
    to confirm these suspects. %

\para{Precision analysis}
In the ideal case, %
every race condition confirmed by \sys{} is real,
    assuming that its model of the NOS under test is accurate and complete.

In real-system deployments, \sys{} does not provide strict soundness or completeness guarantees. 
{It may miss race conditions because the selected test cases and bounded perturbations
    cover only part of the execution space.}
Nevertheless, by confirming race condition suspects through perturbed executions, 
    it produces high-precision race condition reports in practice, 
    as shown in the evaluation (\S\ref{sec:evaluation}).

\para{Severity assessment}
Detected race conditions are submitted to \sys{} users, i.e., testing engineers and developers, for severity assessment.
Severity assessment is primarily based on comparing the behaviors of the original and perturbed test executions.
The severity of each reported race condition is measured by how quickly and to what extent it leads to user-visible abnormal program behavior.
Specifically, we consider three measures of severity:
\begin{itemize}[leftmargin=1em]
    \item \textit{First point of divergence}.
        \sys{} identifies the earliest subsequent behavior 
            at which the original and perturbed executions deviate.
        For \sys{} users, the first point of divergence
            provides a concrete starting point
            for understanding the impact
            of a detected race condition.
        It helps developers
            localize the propagation of the race condition,
            and prioritize severe cases. %
    \item \textit{Divergence in outgoing-message sequences}.
        Analyzing divergence in sequences of outgoing messages
            helps \sys{} users understand
            whether and how the effect of a race condition propagates
            beyond the component where it occurs.
        Since NOS components interact primarily
            through asynchronous messages,
            such divergence shows
            how the race condition has altered the component interactions.
    \item \textit{Divergence in message-handler sequences}.
        Comparing message handler sequences
            across executions
            helps \sys{} users understand
            how a race condition affects
            the subsequent control-plane execution.
        Since NOS components are message-driven,
            the message handler sequences provide
            a high-level view of how components process messages
            during a configuration update task.
        Divergence in these sequences indicates
            how the race condition has affected the execution path within a component. %
\end{itemize}

\subsection{\texorpdfstring{Discussion}{Discussion}} \label{subsec:discussions}

\para{\texorpdfstring{{Relation to causal consistency}}{Relation to causal consistency}}
{The happen-before model in \sys{} captures
    the causal order
    between operations~\cite{lamport1978time, ahamad1995causal}.
    In the analysis of test executions, \sys{} respects the causal order between operations.
    However, ensuring causal consistency is far from sufficient to unearth race-condition bugs.

    {\sys{} further explores the reorderings which are permitted by the causal ordering.}
    \sys{} proactively reverses the order between concurrent operations,
    to check whether the execution result is deterministic. %
    That is, beyond respecting the causal ordering,
    \sys{} uses operations that are not causally ordered
    as the key signal for detecting race-condition bugs.

}

\para{{Failure model}}
{Distributed systems are subject to partial failures,
    such as node crashes and network partitions.
The NOS we study runs on a single network device,
    where failures mostly affect the entire device.
Partial failures, e.g., component crashes and message loss, occur only in rare cases,
    and are covered by dedicated fault-injection tests.
Such failures are extremely rare
    in the configuration-update scenarios we target.
So they are not considered by \sys{}.}

\para{\texorpdfstring{{Design limitations}}{Design limitations}}
The current design of \sys{} incorporates 
several simplifications motivated by
practical considerations.
Specifically, with respect to the monitored objects,  
\sys{} currently focuses on simple crucial variables and still needs to investigate how to handle 
complex data structures and database entries.
With respect to the properties being monitored, \sys{} focuses
on crucial variables and behavior differences 
observed within test executions. 
Extending its observation model to include 
cross-component invariants, 
configuration consistency properties, and 
richer state correlations may expose a broader class of race conditions. 
Moreover, richer perturbation mechanisms, 
such as controlled message reordering or 
runtime support from messaging middleware, 
can improve the success rate of 
constructing race-exposing executions. 
However, enhancing the perturbation mechanism 
requires a broader assessment of its system-wide impact.

The current design and implementation of \sys{} 
represent the first step in a longer-term effort. 
We will leverage empirical data 
collected from deployment (\S \ref{sec:impl}) and evaluation (\S \ref{sec:evaluation})
to guide our efforts in addressing these limitations
and developing the next generation of \sys{}.

%% file: framework-overview.tex
\makebox[\textwidth][c]{%
\begin{tikzpicture}[
    >=Latex,
    stage/.style={
        draw=black,
        rounded corners=1.5pt,
        line width=0.6pt,
        text width=2.10cm,
        minimum height=1.10cm,
        align=center,
        font=\footnotesize\bfseries,
        inner xsep=3pt,
        inner ysep=2pt
    },
    phase/.style={
        draw=black,
        dashed,
        line width=0.6pt,
        inner xsep=0.34cm,
        inner ysep=0.15cm
    },
    arr/.style={->, line width=0.6pt, >=Latex},
    lbl/.style={align=center, font=\footnotesize},
    albl/.style={align=center, font=\scriptsize},
    title/.style={align=center, font=\small\bfseries}
]

\def\dx{2.85}
\node[stage] (comp) at (0, 0) {Component\\Identification\\{\normalfont\scriptsize(\S\ref{subsubsec:interacting-components})}};
\node[stage] (vars) at (\dx, 0) {Crucial Variable\\Identification\\{\normalfont\scriptsize(\S\ref{subsubsec:crucial-variables})}};
\node[stage] (group) at (2*\dx+0.9, 0) {Grouping\\{\normalfont\scriptsize(\S\ref{subsubsec:grouping-perturbations})}};
\node[stage] (strengthen) at (3*\dx+0.9, 0) {Strengthening\\{\normalfont\scriptsize(\S\ref{subsubsec:strengthening-perturbations})}};

\node[lbl, anchor=east] (target) at (-1.75, 0.2) {Target component};
\node[lbl, anchor=east] (tests) at (-1.75, -0.2) {Test cases};
\node[lbl, anchor=west] (reports) at ($(strengthen)+(2.4,0)$) {Reports};

\draw[arr] (target.east) -- (comp.west|-target.east);
\draw[arr] (tests.east) -- (comp.west|-tests.east);
\draw[arr] (comp.east) -- (vars.west);
\draw[arr] (vars.east) --
    node[albl, above=5pt, fill=white, inner sep=1pt, xshift=-1pt, yshift=-3pt] {Preparatory\\execution} 
    (group.west);
\draw[arr] (group.east) -- (strengthen.west);
\draw[arr] (strengthen.east) -- 
    node[albl, above=5pt, fill=white, inner sep=1pt, xshift=-2pt, yshift=-5pt] {Detection} 
    node[albl, below=5pt, fill=white, inner sep=1pt, xshift=-2pt, yshift=5pt] {Validation}
    (reports.west);

\begin{scope}[on background layer]
    \node[phase, fit=(comp)(vars)] (phase1) {};
    \node[phase, fit=(group)(strengthen)] (phase2) {};
\end{scope}
\node[title, anchor=south, inner ysep=0pt, yshift=0.1cm] at (phase1.north) {Phase 1: Concentration};
\node[title, anchor=south, inner ysep=0pt, yshift=0.1cm] at (phase2.north) {Phase 2: Perturbation};
\end{tikzpicture}%
}

%% file: impl.tex
\section{Implementation and Integration} \label{sec:impl}

We first describe the \sys{} implementation,
    followed by its integration into an industrial-grade NOS. %

\para{Implementation}
\sys{} is implemented as a dynamic analysis tool 
    for NOS control planes, 
    centered on runtime tracing and offline analysis.
Runtime tracing relies on Clang-based source instrumentation
    \cite{clangastmatchers}.
The generated hooks in selected NOS components record 
    crucial-variable accesses,
    message exchanges, and 
    the execution context. %
Message sender-side hooks also inject message delays 
    during perturbation executions.

Offline analysis identifies concurrent accesses on crucial variables, derives delay rules for
    perturbation executions, and compares original and perturbed executions
    to detect race conditions.
{Within each execution, it distinguishes runtime instances
    using the source-level variable identifier and object address.
Since object addresses may change across executions, it matches original and
    perturbed accesses using the variable identifier and recorded execution context.}
\sys{} is implemented in
    over 9,000 lines of C/C++ code
    and 25,000 lines of Python code.
Further implementation details are provided
\relax
    in Appendix~\ref{app:impl}.%
\relax

\para{Integration}
\sys{} is integrated into the testing factory 
    of the target NOS for monthly release testing.
The testing factory executes system-level test plans 
    on virtualized network topologies, 
    persistently stores test artifacts, and %
    publishes test results for further analysis.
\relax
{Further details of the testing factory are provided in Appendix~\ref{app:factory}.}
\relax

To integrate \sys{} into the testing factory, an \textit{orchestrator} manages each
    \sys{} run as a dependency graph of test executions and
    offline-analysis steps.
The orchestrator submits test plans to the factory,
    retrieves test artifacts for analysis,
    and submits follow-up plans generated from analysis results.
\relax
The detailed orchestrator workflow of \sys{} is described
    in Appendix~\ref{app:orchestration}.%
\relax

Applying \sys{} to a new target component only requires the component name,
    functional labels for relevant configuration-update tests, and optional
    hints about related components or tasks.
This configuration typically takes less than 30 minutes to prepare
    and is reused in subsequent runs.

%% file: exp.tex
\section{Evaluation} \label{sec:evaluation}

In this section, we evaluate the performance of \sys{}
    with respect to the following questions: 
    1) How effective is \sys{} in detecting race conditions and finding bugs? (\S\ref{subsec:evaluation-effectiveness});
    2) How efficient is \sys{}{, and what limits the effectiveness of perturbation}? (\S\ref{subsec:evaluation-efficiency}).

\input{eval-confirmed-reports}

We evaluate \sys{} on 5 target components
    in a production on-device NOS.
ARP~\cite{rfc826}, DHCP~\cite{rfc2131}, and VRRP~\cite{rfc9568} are
    protocol-derived labels for components that provide
    protocol/application services.
MGR-A and MGR-B are anon\-y\-mized labels for components that provide
    device-state orchestration services.
Each component contains 500k+ lines of code,
    and the testing factory provides 5,000+
    configuration-update test cases per component on average.

For each target component, 
    \sys{} runs 200 preparatory executions.
Each preparatory execution gives rise to 
    3--5 perturbation groups on average, 
    due to conflicts between perturbations.
These groups are submitted as parallel perturbed executions.
For unsuccessful perturbations, 
    \sys{} performs up to 3 rounds of perturbation strengthening.

The experiments mainly evaluate 
the effectiveness and efficiency of \sys{}.
The effectiveness evaluation focuses on three aspects: 
\sys{}'s ability to detect race conditions, 
the severity of the detected race conditions, and 
the rate at which they are confirmed as real system bugs.
The effectiveness evaluation further examines 
how perturbation strengthening affects 
the ability of \sys{} to detect race conditions.
The efficiency evaluation focuses on 
    the overhead incurred by \sys{} 
    in detecting race conditions, and 
    specifically evaluates the contribution 
    of the concentration phase 
    to overhead reduction.

\subsection{Effectiveness} \label{subsec:evaluation-effectiveness}

Across the 5 target components, 
    \sys{} generates 32 race condition reports,
    each accompanied by severity assessment results 
    to support developer review.
Developers confirmed 66\% (21 out of 32) 
    of the reported race conditions as severe bugs.
\sys{} exposed all 21 confirmed bugs using existing tests, 
    even though the same tests had repeatedly passed in the testing factory.
Among the 21 developer-confirmed bugs, 
    15 were previously unknown, and 6 were known bugs 
    which were recently reported and
    were still open at the time.
The severity assessment results 
provide effective guidance for developers to understand 
the downstream impact of the race conditions and 
locate the relevant buggy code fragments.
{Analyzing the 32 reports took about 20 person-days in total,
    0.25\textasciitilde 1 person-day per report,
    mostly by developers responsible for the corresponding components.
For each report, they examined the relevant code
    to understand the cross-component interactions and determined whether the race condition was harmful or benign.}

All 21 confirmed bugs have now been fixed.
Given the effectiveness of \sys{} in this deployment,
    the development team is applying \sys{}
    to more control-plane components in our target NOS,
    as well as to other production NOSes having similar architecture.

\input{eval-effectiveness-overall}
\subsubsection{Developer-confirmed bugs}

\tabref{tab:evaluation-confirmed-reports}
summarizes the 21 bug-triggering race conditions. %
The number of race conditions is similar across the target components 
(4\textasciitilde 5 race conditions per component),
suggesting that the occurrence of bug-triggering race conditions 
is weakly correlated with component-specific functionality and 
more strongly associated with the pattern of 
asynchronous interactions among components.

For each bug-triggering race condition, 
\tabref{tab:evaluation-confirmed-reports} reports 
the associated configuration update task
and the crucial variable involved.
Note that ARP\#1 is the motivating example in \figref{fig:intro-motivation}. %
For each report, the severity assessment highlights 
differences in outgoing-message and 
message-handler sequences 
between the test executions with and without perturbation.
The severity assessment helps developers reason about 
the consequences of the reported race conditions.
As shown in \tabref{tab:evaluation-confirmed-reports},
most confirmed bugs have severe consequences, such as stale control-plane state,
inconsistent device state, forwarding-state errors, and service unavailability.
{Fixing these bugs could also be complex.
A bug may be exposed in one component,
    but the buggy message order and related state changes
    may involve several components.
Fixes therefore ranged from local changes
    to global ones requiring coordination across teams.}

\input{eval-overall-workflow-cost}
\subsubsection{From concurrent accesses to race conditions} \label{subsubsec:evaluation-ca-to-rc}

\tabref{tab:evaluation-effectiveness}
illustrates how \sys{} starts from suspected race conditions, 
    i.e., concurrent accesses, and 
    progressively filters out false positives 
    using perturbed executions. 
The results show a substantial reduction 
    from observed concurrent accesses 
    to the race conditions reported by \sys{}.
Across the 5 components, 
    \sys{} observes over 800 concurrent accesses, 
    with at least 120 in every component.
\sys{} reports 32 race conditions in total, 
    which are 4\% of the original concurrent accesses.

The majority of the false-positive filtering occurs 
    when \sys{} attempts to obtain corresponding perturbed accesses.
Under the initial perturbation, 
    less than 10\% of the concurrent accesses 
    have a corresponding perturbed access.
Perturbation strengthening (\S\ref{subsubsec:strengthening-perturbations}) 
    raises this ratio to about 20\%.
The message delays injected are bounded by 
    the timeout values used in the system implementation.
In practice, the perturbation strengthening 
    runs for at most 3 rounds.
Beyond 3 rounds, larger delays increase 
    the side effects of perturbation, 
    making it harder for \sys{} to observe 
    the original concurrent accesses.

\sys{} further checks whether the values of the crucial variable differ 
    across the two executions.
When multiple related race conditions involve the same crucial variable 
    and have the same pattern of variable access by message handlers,
    \sys{} reports one %
    race condition with
    the corresponding evidence for developer review.
Overall, \sys{} reports 32 race conditions, 
    which are 20\% of the concurrent accesses 
    paired with perturbed accesses.

\subsection{Efficiency and Limitations} \label{subsec:evaluation-efficiency}

{We measure the execution and offline analysis costs of \sys{}.
We also examine how concentration reduces detection cost
    and what limits the effectiveness of perturbation.}
The cost for race condition detection by \sys{} 
    is shown in \tabref{tab:evaluation-overall-cost}.
For each component, \sys{} runs 
    200 preparatory executions, and
    no more than 720 perturbed executions.
These executions are based on 200 of the selected test cases, 
    greedily chosen according to test coverage statistics (as discussed in \S\ref{subsec:concen}).
Individual preparatory and perturbed executions take 
    3\textasciitilde 5 and 
    5\textasciitilde 6 minutes, respectively.
For each component, 
    each preparatory execution derives 40 perturbations on average.
\sys{} partitions them into only 3\textasciitilde 5 mutually conflicting groups,
    which are executed in parallel.
Perturbation planning keeps 
    the execution time for each component 
    within one day,
    without which the experiment time would 
    increase by an estimated factor of 10.

\input{eval-deployable-overhead}
\subsubsection{Cost of enabling \sys{}} \label{subsubsec:evaluation-inflation-rate}

The execution-time and memory overheads of enabling \sys{} tracing
    are shown in \tabref{tab:evaluation-overhead}.
We normalize tracing overhead against baseline executions without enabling \sys{}.

For the 5 components tested, 
    the execution-time and memory overheads are no more than 
    10\% and 20\%, respectively.
We also measure CPU utilization 
    before and after enabling tracing.
Enabling tracing increases CPU utilization
    by no more than 10 percentage points.
Across different overhead metrics, 
    \sys{} introduces only modest execution overhead, 
    making it practical for large-scale deployment 
    in the existing testing environment 
    for race condition detection.

\subsubsection{Cost for trace analysis} \label{subsubsec:evaluation-offline-analysis-cost}

\sys{} performs offline analysis on traces 
    collected from test executions. 
\tabref{tab:evaluation-offline-cost} reports 
    the trace size and 
    analysis time. %
For most target components, the per-execution trace size is below 20 MB,
and offline analysis finishes within one minute for all target components.

We further amortize trace-analysis cost over the race conditions reported by \sys{}.
For each reported race condition, 
    \sys{} usually analyzes hundreds of MB of traces
    and spends minutes in offline analysis.
The longest analysis 
    finishes within 15 minutes.
Compared with the tens of hours spent on 
    preparatory and perturbed executions for each component, 
    the analysis cost is practical for 
    obtaining race condition reports.

\input{eval-offline-analysis-cost}

\subsubsection{Impact of concentration on cost reduction} \label{subsubsec:evaluation-concentration-cost-reduction}

\sys{} reduces the cost of race condition detection 
    along three dimensions: components, variables, and test cases, 
    as discussed in \S\ref{subsec:concen}.
We first quantify, through experiments, 
    the reduction along the above three dimensions, and then 
    analyze how this reduction contributes to 
    lowering the cost of race condition detection.

{As shown in \tabref{tab:evaluation-concentration-results},} the concentration phase selects fewer than 10 interacting components (9\% on average),
    from tens of components involved in the configuration update task.
Concentration selects tens to hundreds of crucial variables (24\% on average),
    from hundreds to more than one thousand variables
    that store component state information.
As for test execution, hundreds of test cases are selected for each target component (8\% on average), 
    from thousands or tens of thousands of related test cases in the testing factory.

The effectiveness of the concentration phase
    is reflected in the substantial cost reduction
    in race condition detection.
Compared with the no-concentration baseline,
    enabling concentration reduces the normalized execution-time and memory overheads
    from 90\% to 10\% and from 65\% to 20\%, respectively.
CPU utilization also decreases from 80\% to 55\%.

\input{eval-concentration-cost}
\subsubsection{Limitations of perturbation} \label{subsubsec:evaluation-perturbation-limitations}

As shown in \tabref{tab:evaluation-effectiveness}
    (\S\ref{subsubsec:evaluation-ca-to-rc}),
\sys{} obtained corresponding perturbed accesses 
    for only 19\% of the concurrent accesses (157 out of 824).
To understand this reduction, 
    we inspected the concurrent accesses 
    not paired with %
    perturbed accesses.
For each of these concurrent accesses, 
    we measured the time span %
    between the configuration update commands. %
The time spans between commands were usually large:
    less than 10\% had a 1\textasciitilde 5 second span, 
    about 25\% had a 5\textasciitilde 10 second span, and
    over 60\% had a span above 10 seconds.

A long time interval between commands limits the effects of message delay injection.
In our perturbation strengthening experiments, 
    delays above about 5 seconds often drove executions 
    into timeout or retry paths.
\sys{} then often failed to observe 
    the same concurrent access occurrence. %
This observation suggests that 
    improving perturbation effectiveness requires test cases 
    that place related configuration commands closer together.
A practical approach is to derive such test cases  
    from existing ones. %
For example, 
    a test case can apply, withdraw, and then re-apply 
    the same configuration command, so that state changes 
    occur within a shorter interval.
This gives \sys{} a better chance to obtain 
    the corresponding perturbed access,
    while the side effects of delay injection remain under control.

\subsection{Discussion}

Our evaluation highlights the strengths and limitations of \sys{}
    and offers lessons for improving race-condition detection and prevention in industrial NOSes.

\subsubsection{Benign race conditions}

In the evaluation, developers 
    classified 6 reported race conditions as benign.
We inspected their feedback 
    to understand why these reported race conditions
    are not classified as bugs.
The first reason is that 
    differences in crucial variable values may not violate
    the correctness invariant of the current configuration update task.
For example, 
    in a VRRP switchover task~\cite{rfc9568},
    several physical routers are configured for the same virtual router %
    with one virtual IP address.
The correctness invariant is that 
    exactly one of them should be the active router
    for that virtual router.
A different handler order may make \sys{} observe a different role value 
    on one router for the same switchover task.
If the group still has exactly one active router, 
    developers classify the report as benign.

The second reason is that 
    some components use periodic synchronization or retry 
    to maintain consistency with upstream components or configuration databases.
In several reports, 
    developers explained that the value difference may cause 
    a transient intermediate abnormal state, but
    later synchronization or retry rewrites the affected variable and
    restores the normal state.
Developers therefore did not treat the report 
    as requiring a fix.

To cope with benign race conditions, a possible improvement is to 
    provide an interface for developers to specify
    system-specific correctness invariants over crucial variables
    during offline analysis.
These invariants can then be used to 
    filter out benign race conditions.

\subsubsection{\texorpdfstring{{Lessons on preventing race-condition bugs}}{Lessons on preventing race-condition bugs}}

{A configuration task consists of multiple commands,
    which may read and write shared resources concurrently.
Processing a single command may involve asynchronous message exchanges
    between components and multiple updates to the shared resources.
Developers typically expect these updates to be atomic%
,
    and concurrent commands to be isolated from one another.}

{One possible solution is to store shared resource states
    in a transactional database.
For each command, updates to shared resources
    can be grouped into a single transaction.
Atomicity makes these updates all-or-nothing,
    while isolation prevents concurrent transactions from interfering with one another.
However, transactions can impose substantial runtime overhead.}

{The NOS we study instead synchronizes shared resource state periodically
    and retries failed resource accesses
    to resolve conflicts between concurrent commands.
These mechanisms achieve low runtime overhead,
    at the cost that deeply hidden race-condition bugs still occur.
\sys{} helps uncover such deep bugs.}

%% file: eval-confirmed-reports.tex
\begin{table*}[!t]
\centering
\caption{\textbf{Developer-confirmed bugs from race condition reports.}
``Severity Assessment'' shows outgoing-message or handler-sequence differences.
``Consequence'' shows the functional consequence identified by developers.}
\label{tab:evaluation-confirmed-reports}
\vspace{-8pt}
\small
\setlength{\tabcolsep}{3pt}
\renewcommand{\arraystretch}{1.02}
\begin{tabular*}{\textwidth}{@{\extracolsep{\fill}}llllll@{}}
\toprule
Bug ID &
Task &
Crucial Variable &
Severity Assessment &
Consequence &
Status \\
\midrule
ARP\#1 & VLAN del/add &
subscription state &
add request skipped &
interface unsubscribed &
new/fixed \\
ARP\#2 & ARP-entry cleanup &
delete flag &
entry cleanup skipped &
obsolete ARP entry &
new/fixed \\
ARP\#3 & VLAN-type cleanup &
VLAN type &
deletion notice incomplete &
outdated forwarding entry &
new/fixed \\
ARP\#4 & VLAN group delete &
VLAN group state &
peer ARP update missed &
peer ARP table outdated &
new/fixed \\
ARP\#5 & host tracking disable &
host-tracking state &
route update suppressed &
host route missing &
known/fixed \\
DHCP\#1 & DHCP relay enable &
relay service state &
client request ignored &
client address missing &
new/fixed \\
DHCP\#2 & session-option update &
session option state &
transient option sent &
session setup failed &
new/fixed \\
DHCP\#3 & virtual-router delete &
server service state &
unsubscribe skipped &
subscriptions remain &
new/fixed \\
DHCP\#4 & service/data registration &
service type &
data sent without service &
service record missing &
known/fixed \\
VRRP\#1 & VRRP-entry delete &
node priority &
peer update suppressed &
peer priority outdated &
new/fixed \\
VRRP\#2 & interface bind/unbind &
interface binding &
unbind skipped &
interface binding mismatch &
new/fixed \\
VRRP\#3 & VRRP6 del/add &
node role &
initialization skipped &
route unavailable &
known/fixed \\
VRRP\#4 & VRRP failover &
failover delay state &
delay exit missed &
failover delayed &
known/fixed \\
MGR-A\#1 & user auth enable &
auth status &
config update missed &
stale auth state &
new/fixed \\
MGR-A\#2 & next-hop update &
next-hop selection &
invalid next-hop retained &
route uses invalid next hop &
new/fixed \\
MGR-A\#3 & service registration &
provider state &
early request ignored &
service availability gap &
new/fixed \\
MGR-A\#4 & DHCP relay disable &
relay subscription state &
unsubscribe skipped &
relay subscription leaked &
new/fixed \\
MGR-B\#1 & redundancy update &
node validity state &
valid-state update missed &
failover unavailable &
new/fixed \\
MGR-B\#2 & service-type registration &
ARP service type &
registration skipped &
ARP service unavailable &
new/fixed \\
MGR-B\#3 & EVPN config rollback &
EVPN service-binding state &
cleanup skipped &
IP forwarding broken &
known/fixed \\
MGR-B\#4 & link-aggregation update &
link membership state &
stale member used &
traffic on removed port &
known/fixed \\
\bottomrule
\end{tabular*}%
\par\vspace{-0.2cm}
\end{table*}

%% file: eval-effectiveness-overall.tex
\begin{table}[t]
\centering
\caption{\textbf{Filtering race condition suspects.}
CAs are deduplicated concurrent accesses;
PAs are CAs with corresponding perturbed accesses.
``Init.'' uses the initial perturbations;
``Final'' includes bounded perturbation strengthening.
Percentages are computed over \# CAs in each row.}
\label{tab:evaluation-effectiveness}
\vspace{-8pt}
\small
\begin{tabular*}{\columnwidth}{@{\extracolsep{\fill}}lrrrr@{}}   
\toprule
Target &
\# CAs &
\# Init. PAs &
\# Final PAs &
\# Reports \\
\midrule
ARP   & 173 & 21 (12\%) & 38 (22\%) & 8 (5\%) \\
DHCP  & 127 & 8 (6\%)   & 18 (14\%) & 5 (4\%) \\
VRRP  & 154 & 16 (10\%) & 27 (18\%) & 6 (4\%) \\
MGR-A & 231 & 19 (8\%)  & 49 (21\%) & 7 (3\%) \\
MGR-B & 139 & 11 (8\%)  & 25 (18\%) & 6 (4\%) \\
\midrule
Total & 824 & 75 (9\%)  & 157 (19\%) & 32 (4\%) \\
\bottomrule
\end{tabular*}%
\par\vspace{-0.2cm}
\end{table}

%% file: eval-overall-workflow-cost.tex
\begin{table}[t]
\centering
\caption{\textbf{Execution cost for race condition detection.}
``Total Time'' is the cumulative runtime across executions.}
\label{tab:evaluation-overall-cost}
\vspace{-8pt}
\small
\begin{tabular*}{\columnwidth}{@{\extracolsep{\fill}}lrrrr@{}}
\toprule
Target &
\multicolumn{2}{c}{Preparatory} &
\multicolumn{2}{c}{Perturbed} \\
\cmidrule{2-3}\cmidrule{4-5}
&
\#Execs. &
Total Time &
\#Execs. &
Total Time \\
\midrule
ARP & 200 & 16.70h & 711 & 70.14h \\
DHCP & 200 & 9.87h & 686 & 56.91h \\
VRRP & 200 & 14.00h & 716 & 57.18h \\
MGR-A & 200 & 12.45h & 710 & 62.77h \\
MGR-B & 200 & 12.17h & 651 & 67.02h \\
\midrule
Total & 1000 & 65.19h & 3474 & 314.02h \\
\bottomrule
\end{tabular*}%
\par\vspace{-0.4cm}
\end{table}

%% file: eval-deployable-overhead.tex
\begin{table}[t]
\centering
\caption{\textbf{Overhead of enabling \sys{} tracing.}
Baseline disables tracing on the same executions.
Parentheses show relative time/memory changes.
CPU utilization changes are in percentage points.}
\label{tab:evaluation-overhead}
\vspace{-8pt}
\small
\begin{tabular*}{\columnwidth}{@{\extracolsep{\fill}}lrrr@{}}
\toprule
Target &
Time $\Delta$ &
Mem. $\Delta$ &
CPU $\Delta$ \\
\midrule
ARP & +6.8s (+2.8\%) & +543MiB (+6.3\%) & +0.3 \\
DHCP & +1.4s (+0.6\%) & +334MiB (+3.9\%) & +0.5 \\
VRRP & +13.9s (+7.4\%) & +1397MiB (+16.6\%) & +1.7 \\
MGR-A & +22.2s (+9.5\%) & +1080MiB (+12.8\%) & +7.4 \\
MGR-B & +2.1s (+0.8\%) & +1005MiB (+11.9\%) & +7.5 \\
\midrule
Median & +6.8s (+2.8\%) & +1005MiB (+11.9\%) & +1.7 \\
\bottomrule
\end{tabular*}%
\par\vspace{-0.2cm}
\end{table}

%% file: eval-offline-analysis-cost.tex
\begin{table}[t]
\centering
\caption{\textbf{Offline analysis cost.}
Per-report values amortize total trace size and analysis time.}
\label{tab:evaluation-offline-cost}
\vspace{-8pt}
\small
\begin{tabular*}{\columnwidth}{@{\extracolsep{\fill}}lrrrr@{}}
\toprule
Target &
\multicolumn{2}{c}{Per Execution} &
\multicolumn{2}{c}{Per Report} \\
\cmidrule{2-3}\cmidrule{4-5}
&
Trace &
Time &
Trace &
Time \\
\midrule
ARP & 6.0MB & 23.5s & 120.8MB & 7.9min \\
DHCP & 79.1MB & 14.9s & 2736.9MB & 8.6min \\
VRRP & 18.8MB & 28.0s & 470.0MB & 11.7min \\
MGR-A & 10.4MB & 22.5s & 117.4MB & 4.2min \\
MGR-B & 3.7MB & 13.4s & 91.3MB & 5.5min \\
\midrule
Median & 10.4MB & 22.5s & 120.8MB & 7.9min \\
\bottomrule
\end{tabular*}%
\par\vspace{-0.2cm}
\end{table}

%% file: eval-concentration-cost.tex
\begin{table}[t]
\centering
\caption{\textbf{Scale reduction from concentration.}
Each cell reports before$\to$after concentration counts.}
\label{tab:evaluation-concentration-results}
\vspace{-8pt}
\small
\begin{tabular*}{\columnwidth}{@{\extracolsep{\fill}}lrrr@{}}
\toprule
Target &
Components &
Variables &
Tests \\
\midrule

ARP & $80\to4$ & $1489\to412$ & $8650\to356$ \\
DHCP & $74\to5$ & $790\to125$ & $5219\to261$ \\
VRRP & $82\to8$ & $665\to183$ & $1041\to204$ \\
MGR-A & $79\to7$ & $1628\to415$ & $16198\to448$ \\
MGR-B & $45\to6$ & $419\to94$ & $5206\to405$ \\
\midrule
Median & $79\to6$ & $790\to183$ & $5219\to356$ \\
\bottomrule
\end{tabular*}%
\par\vspace{-0.3cm}
\end{table}

%% file: rw.tex
\section{Related Work} \label{sec:related}

Existing testing techniques have 
    limited effectiveness in race condition detection 
    for concurrent and distributed systems.
In particular, system-level stress testing relies on 
    non-deterministic scheduling and therefore 
    cannot systematically explore rare message interleavings \cite{burckhardt2010pct,leesatapornwongsa2016samc,musuvathi2008chess}.
Formal methods establish rigorous correctness guarantees
    through theorem proving or model checking 
    under a formal specification \cite{lamport2002tla,hawblitzel2015ironfleet,wilcox2015verdi,sharma2023grove,li2023timepiece,ouyang2025multi}.
However, in industrial NOSes, 
    formal verification is often impractical, 
    due to its high learning curve and substantial adoption cost \cite{garavel2020expert,brooker2025systems,huang2026lessons}.

To address these limitations,
    prior work has explored both static and dynamic approaches 
    for detecting race conditions 
    in concurrent and distributed systems.
Static analysis techniques are widely used 
    to detect race conditions in concurrent programs, 
    often based on lockset analysis or happens-before inference \cite{engler2003racerx,voung2007relay,blackshear2018racerd}.
Such techniques are further extended for race condition detection 
    in asynchronous event-driven programs \cite{hu2018static,liu2021threads}.
However, production NOS control planes 
    consist of loosely coupled components
    that communicate through asynchronous messages.
Existing static analysis techniques 
    must conservatively over-approximate
    possible message interleavings and execution paths, 
    leading to combinatorial path explosion \cite{liu2021threads,cai2021canary}.
This severely limits their scalability on production NOSes with
    million-line-scale codebases.
Static analysis techniques also suffer from limited precision.
Recent studies show that static detectors on large system codebases
    still produce substantial false positives that require manual validation
    \cite{kasikci2012portend,li2024lrminer,holter2025sound}.
Production NOSes widely employ 
    customized optimizations, 
    such as priority message queues and batched state commits, 
    which further hinder existing static analysis techniques from 
    precisely detecting race conditions.
\sys{} uses static analysis primarily to identify crucial variables and
    message-interaction perturbation points. 
Race condition suspects are identified
    mainly from dynamic execution information and 
    confirmed through perturbed executions.

Dynamic analysis techniques improve the precision of race condition 
    detection by observing concrete runtime behaviors and execution states \cite{savage1997eraser,flanagan2009fasttrack,serebryany2009threadsanitizer,raychev2013effective,roemer2020smarttrack,kulkarni2021dynamic,zhang2025efficient}.
However, applying dynamic analysis to production NOSes remains challenging 
    due to the high cost of executing complex distributed control-plane workloads 
    and the enormous space of possible message interleavings.
Deterministic simulation and systematic exploration techniques 
have been proposed to improve interleaving coverage
in concurrent and distributed systems \cite{flanagan2005dpor,yang2009modist,leesatapornwongsa2016samc,deligiannis2023coyote,tang2024sandtable,pick2023psym}.
However, exhaustive deterministic control incurs substantial runtime and engineering overhead, 
making such approaches difficult to deploy in industry-scale NOSes.
In contrast, \sys{} employs probabilistic delay injection 
at carefully selected execution 
points to efficiently expose race condition-prone asynchronous interleavings 
while maintaining low execution overhead.
Existing distributed concurrency-bug detectors 
    identify suspicious accesses or message orderings 
    from recorded executions, and validate them 
    through runtime triggering \cite{liu2017dcatch,lu2018cloudraid,lu2022cloudraid}.
Recent work further uses static analysis to guide test generation 
    for such bugs in RPC-based distributed systems \cite{cao2026themis}.
In both cases, bug validation largely depends on visible failure symptoms.
In NOS configuration update tasks, however,
    race-condition-induced bugs often corrupt control-plane state silently
    \cite{tang2021campion,lou2022oathkeeper,lou2025t2c}.
\sys{} therefore makes this silent corruption observable to developers
    by reporting crucial-variable value differences and subsequent behavioral differences.

%% file: conclusion.tex
\section{Conclusion} \label{sec:conclusion}

In this work, we present \sys{}, 
    a dynamic analysis framework for race condition detection in on-device NOSes.
\sys{} tackles the problem with a two-phase design. 
The concentration phase reduces the analysis cost by 
    exploiting the locality of configuration update tasks, while 
    the perturbation phase proactively perturbs the message asynchrony to 
    confirm race condition suspects.
Integrated into the existing testing factory of a production NOS, \sys{} demonstrates that dynamic race condition detection can be made practical for large-scale actively developed NOSes.

Several directions remain 
    for future work. 
One direction is to support incremental analysis as the NOS codebase evolves.
Future work could reuse traces and analysis results across code revisions
    to update race-condition candidates incrementally, reducing the cost of
    deploying \sys{} in CI pipelines.
Another direction is to 
    enrich severity analysis. 
\sys{} currently analyzes how race conditions 
    affect subsequent message-sending and 
    function-call sequences. 
Future work could further incorporate 
    configuration consistency checks, 
    cross-component state invariants, and 
    user-visible service-level symptoms, 
    to help developers prioritize detected race conditions.
Finally, it would be valuable to study whether 
    \sys{}'s two-phase design %
    generalizes to other asynchronous multi-component systems. 
Production distributed control systems
    and cloud management platforms
    share similar challenges: 
    asynchronous message interactions,
    large execution spaces, and 
    limited testing budgets.

%% file: app-testing-factory.tex
\section{Testing Factory of the Target NOS} \label{app:factory}

The testing factory of the target NOS is a large-scale
  Continuous Integration (CI) platform that supports NOS development and release testing.
The factory executes system-level test cases
  on virtualized multi-device topologies.
These test cases configure devices through Command Line Interface (CLI) commands
  and inject test traffic to validate NOS functionality, performance, 
  and stability.
A large portion of them are configuration-update test cases,
    which apply configuration commands and validate the resulting NOS behavior.
\sys{} reuses these configuration-update test cases and their profiles,
  together with the factory's execution environment.

\subsection{Test Execution Environment}
\label{app:test-exec-env}

System-level test executions are specified in \textit{test plans}
    submitted to the factory.
These plans may come from CI jobs, release-test campaigns, or developer
    debugging sessions.
The factory schedules these plans across a pool of virtualized testbeds,
    tracks the state of each execution, and stores the artifacts produced by
    completed executions.
The testbed pool contains over 10{,}000
    virtualized testbeds, and the factory keeps over 5{,}000
    test executions running concurrently.
The factory sustains a daily throughput of over 200{,}000
    system-level test executions.

Each test plan follows a factory-defined schema that specifies the topology to
    instantiate, the workload to execute, and the checks used to determine the
    test outcome.
The topology specification defines the nodes and links in the test:
    Device Under Test (DUT) nodes represent network devices running the NOS under test;
    peer nodes provide neighboring devices along the packet forwarding paths; %
    traffic endpoints inject or observe test traffic;
    and virtual links connect the nodes.
The workload specifies the CLI commands to issue, 
    the test traffic to inject or observe, and 
    the device-state queries to run. %
The checks validate device-local outcomes and cross-device behaviors,
    such as reachability, forwarding behavior, convergence, 
    and state consistency.
A test plan can also include auxiliary
    files and execution-specific parameters. %

The \textit{factory manager} schedules test plans across the testbed pool.
It maintains queues of pending plans, tracks which testbeds are available or
    occupied, and chooses a compatible testbed for each plan based on the
    plan's topology requirements.
The manager keeps multiple plans executing in parallel by assigning queued plans
    to testbeds as they become available.
When multiple plans use the same or similar topology, the manager can place
    them on suitable testbeds to reduce redundant topology setup.

The testbed pool supplies the virtualized environments
    used for test executions.
A \textit{testbed manager} controls each testbed.
Once the factory manager assigns a plan to a testbed, 
    the testbed manager sets up the requested topology. %
KVM-based virtual machines run NOS device instances for DUT nodes and for peer
    nodes that require a full NOS image.
Docker containers run traffic endpoints and auxiliary test services.
The testbed manager connects these devices and services with virtual links,
    producing the multi-device NOS environment specified by the test plan.

Once the environment is ready,
    the testbed manager executes the workload
    using the auxiliary files and parameters specified in the test plan,
    and evaluates the checks to determine the test outcome.

At the end of a test execution, the testbed manager writes command outputs, 
    device logs, traffic observations, state snapshots, and test results 
    to the \textit{artifact store} of the testing factory, 
    which is a storage service backed by file systems and databases.
The artifact store indexes these artifacts by test execution information and 
    records each artifact's type and storage location, 
    allowing downstream tools to retrieve the artifacts 
    for a given test execution.
{The testbed is then released back to the pool for later test executions.}

The factory also includes a unified reporting system.
It presents the factory's standard test outcomes together with analysis reports
    generated by external tools, and links each report to 
    the corresponding test plans, executions, and artifacts.
This enables test engineers to inspect both execution outcomes 
    and analysis results within the same testing workflow.

\subsection{Configuration-Update Test Cases}
\label{app:config-update-tests}

The factory maintains more than 200,000 configuration-update test cases.
These test cases exercise ordered sequences of configuration commands issued to
  one or more DUTs in the test topology.
The covered update scenarios arise from operational changes, including
  topology changes, traffic dynamics, and policy revisions.
These updates can affect connectivity, protocol behavior, forwarding
  state, or access policy.
Each test case captures the lifecycle of a configuration update.
It first establishes the pre-update device state and then applies one or more
  ordered configuration changes.
It checks outcomes both
  during and after the update.
Together, these test cases span hundreds of NOS components and thousands of configuration features.
Thousands of configuration-update test cases are added or updated
  in each monthly NOS release cycle.

The factory maintains a profile for each configuration-update test case to organize
  and select test cases at scale.
Each profile contains functional labels and coverage information.
The functional labels classify each test case by the NOS function, protocol or
  service category, and configuration behavior it is intended to cover.
{Test engineers assign and maintain these labels as test-case metadata.}
The factory collects coverage information through a dedicated profiling
pipeline.
The pipeline periodically runs test cases on NOS images built with coverage instrumentation
  and records the results in their profiles.
Figure~\ref{fig:test-profile} shows a simplified, anonymized configuration-update test case and its profile.

\begin{figure}[t]
\centering
\begin{minipage}{0.98\columnwidth}
\begin{lstlisting}[style=profileyaml]
test_case: arp_neighbor_refresh
topology:
  class: 3-DUT L3 ring
  duts: [dut-a, dut-b, dut-c]
  traffic_endpoints: [tg0]
  virtual_links: 4
workload:
  update_scope: multi-DUT interface-subnet update
  steps:
    - apply subnet changes on dut-a and dut-b
    - inject ARP/IP traffic through tg0
    - query ARP table and interface state on DUTs
    - validate neighbor convergence and reachability
profile:
  last_profiled: R26.05
  functional_labels:
    nos_function: ARP
    service_category: neighbor discovery
    config_behavior: interface-subnet update
  execution_history:
    recent_runs: 128
    median_duration: 7.4 min
  function_coverage:
    ARP: [refresh_neighbor, resolve_nexthop]
    MGR-A: [apply_intf_cfg, publish_state]
\end{lstlisting}
\end{minipage}
\vspace{-0.45em}
\caption{A simplified and anonymized configuration-update test case and its profile.}
\Description{A YAML-like excerpt of a configuration-update test case and its profile, showing topology, workload steps, functional labels, execution history, and function coverage.}
\label{fig:test-profile}
\end{figure}

\sys{} uses functional labels to identify test cases related to each target component.
In our evaluation, the label-based candidate set contains more than 5,000
  configuration-update test cases per target component on average.
\sys{} then prioritizes test cases that cover functions containing crucial-variable access sites,
  based on the function-level coverage in these profiles (\S\ref{subsubsec:preparation}).

Overall, the integration of \sys{} relies on testing factory support for setting up test
  topologies, scheduling test executions, retrieving test case profiles and execution artifacts,
  and presenting analysis reports.
These are common capabilities of production NOS CI infrastructures.
An NOS CI environment with equivalent support can serve as the execution backend
  for \sys{}.

%% file: app-orchestration.tex
\section{\sys{} Workflow Orchestration}
\label{app:orchestration}

\sys{} builds its end-to-end workflow on the testing factory.
The orchestrator organizes this workflow into
    a dependency graph of test executions and offline-analysis steps.
It tracks the state of each step and submits test plans
    when their {required} inputs are ready.
After each execution, it retrieves
    the resulting artifacts for offline analysis
    and submits follow-up plans based on the analysis results.

Given the target component and functional labels
    for relevant configuration-update test cases,
    the orchestrator first invokes the build pipeline to insert message-passing
    hooks in all NOS components.
It then selects these test cases
    from the testing factory and submits the corresponding test plans.
The factory executes these plans on the instrumented NOS image, and the orchestrator retrieves the message traces for concentration.
After concentration determines the interacting components, crucial variables, and
    selected test cases, the orchestrator invokes the build pipeline again.
The resulting image retains message-passing hooks only in the target and
    interacting components and adds variable-access hooks in the target component.
The orchestrator then submits the test plans
    of the selected test cases for preparatory executions, which produce the traces
    needed by offline analysis.

Offline analysis processes preparatory traces and organizes the resulting
    perturbations into groups.
For each group, the orchestrator creates a perturbation test plan, attaches the
    message-delay rules generated for that group, and submits the plan to the
    factory.
After the factory completes the perturbed executions, the orchestrator retrieves the
    traces for offline analysis.
When offline analysis determines that an unsuccessful perturbation needs strengthening, the orchestrator submits a
    follow-up perturbation plan with increased message delays.
Finally, offline analysis produces race-condition reports, and the orchestrator
    submits these reports to
    the testing factory's reporting system.

The orchestrator records which preparatory execution each perturbation group belongs
    to and which perturbed executions were launched for that group.
These records let offline analysis associate perturbed traces
    with the corresponding preparatory traces, and let the reports
    link each detected race condition to the artifacts of these executions.

%% file: app-implementation.tex
\section{\sys{} Implementation} \label{app:impl}

The implementation details below address two issues central to \sys{}.
Runtime tracing must capture enough context around 
    crucial-variable accesses and message exchanges 
    for ordering and value analysis, 
    while also supporting message-delay perturbation.
Offline analysis must match
    original and perturbed access occurrences across executions 
    even when variable instances and recorded traces differ across runs.

\subsection{Runtime Tracing} \label{app:runtime-tracing}

\sys{} uses Clang-based AST analysis to
    perform source-level instrumentation \cite{clangastmatchers}.
Concentration provides the interacting components and
    crucial variables to be instrumented.
The instrumentation inserts message-passing hooks in the target component and its
    interacting components, and inserts variable-access hooks at
    crucial-variable access sites in the target component.
The generated hook code is compiled into
    the corresponding NOS components.
During system execution, the hook code invokes
    the runtime library to generate execution traces.

Variable-access hooks are placed at read/update access sites of crucial
    variables in the target component.
These accesses may appear inside conditional expressions, loop conditions,
    function arguments, or compound update expressions.
The insertion point therefore determines when the hook is triggered and which
    value is recorded.
\sys{} uses Clang AST to locate these access sites, classify each access
    as a read or update, and choose an insertion point based on the surrounding
    statement.
Specifically, hooks for conditional and loop expressions are placed after the
    expression is evaluated on the executed path, while hooks for update
    expressions record the updated value.
When a variable-access hook is triggered, the runtime library records the
    source-level variable identifier, object address, access type, access value,
    and runtime call stack.

The NOS communication layer provides a small set of common APIs for exchanging
    messages between components.
\sys{} wraps these send/receive API calls with message-passing hooks.
When a message-passing hook is triggered, the runtime library records the
    sender, receiver, message type, and runtime call stack.
{For send/receive pairing within an execution, it also records hashes
    of the complete messages.}
The runtime library records a message operation only when both the sender and the
    receiver are either the target component or one of its interacting
    components.
When a message-receiving hook is triggered, the runtime library checks whether the
    received message should be recorded.
If the message is recorded, the runtime library also records traces from
    the message handler that processes the message and associates them with the recorded
    incoming message.
Offline analysis uses this association to link each variable-access operation to the
    message handler that contains it.

For perturbation executions,
    the same runtime library reads the delay rules.
Before invoking the original send API, the send-API wrapper checks each
    outgoing message and injects the delay specified in the delay rules.

\subsection{Offline Analysis}
\label{app:offline-analysis}

Offline analysis first handles runtime variable instances within each
    execution.
A source-level crucial variable may correspond to multiple runtime objects in
    the same execution.
Offline analysis uses the source-level variable identifier and object
    address to distinguish these instances.
{During the deployment of \sys{}, we observed that only a small fraction of
    these instances were also updated by threads
    other than the component's message-handling thread.
Offline analysis therefore simply excludes them from race condition detection.}

Offline analysis then checks whether each concurrent access occurrence has a
    matching perturbed access occurrence across executions.
This matching does not use object addresses, which may change across executions.
A perturbed access occurrence is matched only when it has the same recorded
    crucial-variable identifier as the original occurrence and
    the two corresponding handler executions
    appear in the reversed order.
Offline analysis
matches handler executions
    using the recorded header fields of the messages
    leading to each handler execution, together with the runtime call stack of the access.
The same matching rule is also used to merge duplicate concurrent and perturbed
    access occurrences, avoiding duplicate race condition reports.

Race condition detection compares the observed values in an original occurrence
    and its matched perturbed occurrence.
Before this comparison, offline analysis checks the recorded access sequences on
    both sides.
For each side, the check scans the recorded access sequence of the same runtime
    variable instance within the configuration update task.
Each observed value change must have a corresponding recorded update access.
The runtime library records traces only for handlers whose incoming messages
    are recorded.
Updates performed while processing other messages may be absent from the trace.
Such missing updates prevent offline analysis from reliably attributing an
    observed value difference to the reversal of handler order.

For each access in the original and perturbed occurrences,
    offline analysis identifies the message handler that contains the access.
It then follows recorded send/receive relations backward from that handler to
    identify the configuration update task to which the handler belongs.
Within that task, offline analysis extracts the recorded access sequence for the
    same runtime variable instance.
If a recorded access has a value different from the previous recorded access
    but is not an update access that records the new value,
    the update that produced the change is missing from the trace.
If either occurrence contains such a missing update,
    \sys{} excludes the matched pair from race condition detection.
Other concurrent or perturbed occurrences are still analyzed.